\documentclass [12pt,eqsecnum,amsfonts,aps]{revtex4}

\input epsf
\newcommand{\beq}{\begin{equation}}
\newcommand{\eeq}{\end{equation}}
\newcommand{\beqs}{\begin{eqnarray}}
\newcommand{\eeqs}{\end{eqnarray}}

\begin{document}

\baselineskip 6.0mm

\title{Weighted Graph Colorings} 

\bigskip

\author{Shu-Chiuan Chang$^{a}$}
\email{scchang@mail.ncku.edu.tw}

\author{Robert Shrock$^{b}$}
\email{robert.shrock@stonybrook.edu}

\affiliation{(a) \ Department of Physics \\
National Cheng Kung University \\
Tainan 70101, Taiwan}

\affiliation{(b) \ C. N. Yang Institute for Theoretical Physics \\
Stony Brook University  \\
Stony Brook, N. Y. 11794}


\begin{abstract}

We study two weighted graph coloring problems, in which one assigns $q$ colors
to the vertices of a graph such that adjacent vertices have different colors,
with a vertex weighting $w$ that either disfavors or favors a given color. We
exhibit a weighted chromatic polynomial $Ph(G,q,w)$ associated with this
problem that generalizes the chromatic polynomial $P(G,q)$. General properties
of this polynomial are proved, and illustrative calculations for various
families of graphs are presented.  We show that the weighted chromatic
polynomial is able to distinguish between certain graphs that yield the same
chromatic polynomial.  We give a general structural formula for $Ph(G,q,w)$ for
lattice strip graphs $G$ with periodic longitudinal boundary conditions. The
zeros of $Ph(G,q,w)$ in the $q$ and $w$ planes and their accumulation sets in
the limit of infinitely many vertices of $G$ are analyzed.  Finally, some
related weighted graph coloring problems are mentioned.

\end{abstract}

\maketitle


\pagestyle{plain}
\pagenumbering{arabic}

\section{Introduction}

Recently we have formulated two weighted graph coloring problems in which one
assigns $q$ colors to the vertices of a graph such that adjacent vertices
(i.e., vertices connected by an edge of the graph) have different colors, with
a vertex weighting $w$ that either disfavors (for $0 \le w < 1$) or favors (for
$w > 1$) a given color \cite{hl}.  We label these with the abbreviations DFCP
and FCP for \underline{d}is\underline{f}avored-\underline{c}olor and
\underline{f}avored-\underline{c}olor weighted graph vertex coloring
\underline{p}roblems.  Since all of the colors are, {\it a priori}, equivalent,
it does not matter which color one takes to be subject to the weighting. In the
present paper we study these interesting weighted graph coloring problems in
detail. An assignment of $q$ colors to the vertices of a graph $G$, such that
adjacent vertices have different colors, is called a ``proper $q$-coloring'' of
the vertices of $G$.  We analyze the properties of an associated weighted
chromatic polynomial that we denote $Ph(G,q,w)$, which generalizes the
chromatic polynomial $P(G,q)$ and constitutes a $w$-dependent measure, extended
from the integers to the real numbers, of the number of proper $q$-colorings of
the vertices of $G$.  In the weighted graph coloring problem, with $q \in
{\mathbb N}_+$ being the number of colors, for a given graph $G$, $P(G,q)$ is a
map from ${\mathbb N}_+$ to ${\mathbb N}$, while $Ph(G,q,w)$ is a map from
${\mathbb N}_+ \times I$ to ${\mathbb R}$, where $I$ denotes the DFCP interval
$0 \le w < 1$ or the FCP interval $w > 1$.  In both cases, one can formally
extend the domain of $q$ and $w$ to ${\mathbb R}$ or, indeed, ${\mathbb C}$,
and the latter extension is, in fact, necessary when one analyzes the zeros of
$P(G,q)$ or $Ph(G,q,w)$, respectively.  The polynomial $Ph(G,q,w)$ is
equivalent to the partition function of the $q$-state Potts antiferromagnet on
the graph $G$ in an external magnetic field $H$, in the limit where the
spin-spin exchange coupling becomes infinitely strong, so that the only spin
configurations contributing to this partition function are those for which
spins on adjacent vertices are different \cite{hl,zth}. There has been
continuing interest in the Potts model and chromatic and Tutte polynomials for
many years; reviews of the Potts model include \cite{wurev}-\cite{wubook} and
reviews of chromatic and Tutte polynomials include \cite{rrev}-\cite{jemrev}.

There are a number of motivations for this study.  One is the intrinsic
mathematical interest of these two new kinds of graph coloring problems.  A
second stems from the equivalence to the statistical mechanics of the Potts
antiferromagnet in a magnetic field. A third is the fact that these weighted
graph coloring problems have physical applications.  For example, the weighted
graph coloring problem with $0 \le w < 1$ (i.e., the DFCP) describes, among
other things, the assignment of frequencies to commercial radio broadcasting or
wireless mobile communication transmitters in an area such that (i) adjacent
stations must use different frequencies to avoid interference and (ii) stations
prefer to avoid transmitting on one particular frequency, e.g., because it is
used for data-taking by a nearby radio astronomy antenna \cite{nrqz}. The
weighted graph coloring problem with $w > 1$ (i.e., the FCP) describes this
frequency assignment process with a preference for one of the $q$ frequencies,
e.g., because it is most free of interference.

We remark on some special cases of $Ph(G,q,w)$.  The case $w=1$ is equivalent
to the usual (unweighted) chromatic polynomial, $P(G,q)$, counting the number
of proper $q$-colorings of the vertices of $G$: 
\beq
Ph(G,q,1) = P(G,q) \ . 
\label{phw1}
\eeq

The chromatic number of $G$, denoted $\chi(G)$, is the minimal number of colors
for which one can carry out a proper $q$-coloring of the vertices of $G$.
For $w = 0$, one is complete forbidden from assigning the disfavored color to
any of the vertices, so that the problem reduces to that of a proper coloring
of the vertices of $G$ with $q-1$ colors without any weighting among these
$q-1$ colors, which is thus described by the usual unweighted chromatic
polynomial $P(G,q-1)$:
\beq
Ph(G,q,0) = P(G,q-1) \ . 
\label{phw0}
\eeq
Thus, the DFCP, described by $Ph(G,q,w)$ with $0 \le w \le 1$, may be regarded
as interpolating between $P(G,q)$ and $P(G,q-1)$.  In the FCP, as $w$ increases
above 1 to large positive values, the favored weighted of one color is
increasingly in conflict with the strict constraint that no two adjacent
vertices have the same color.  Hence, the FCP involves frustration in the
technical sense of statistical mechanics, i.e. mutually conflicting tendencies
built into the system.

\section{Definitions and Some Basic Properties}

\subsection{Relation with Potts Model in an External Magnetic Field} 

Consider a graph $G=(V,E)$, defined by its set of vertices $V$ and edges (=
bonds) $E$. A spanning subgraph $G' \subseteq G$ is defined as the subgraph
containing the same set of vertices $V$ and a subset of the edges; $G' =
(V,E')$ with $E' \subseteq E$. For a graph $G$ we denote the number of
vertices, edges, and connected components as $n(G)$, $e(G)$, and $k(G)$,
respectively.  Where no confusion can result, we shall often abbreviate $n(G)$
as simply $n$. We further denote the connected subgraphs of a spanning subgraph
$G'$ as $G'_i$, $i=1,..,k(G')$.  To obtain an expression for $Ph(G,q,w)$, we
make use of the fact that it is a special case of the partition function for
the $q$-state Potts model in an external magnetic field in the limit of
infinitely strong antiferromagnetic spin-spin coupling.  In thermal equilibrium
at temperature $T$, the general Potts model partition function is given by
\beq
Z = \sum_{ \{ \sigma_n \} } e^{-\beta {\cal H}} 
\label{z}
\eeq
with the Hamiltonian
\beq
{\cal H} = -J \sum_{\langle i j \rangle} \delta_{\sigma_i, \sigma_j}
- H \sum_\ell \delta_{\sigma_\ell,1} \ , 
\label{ham}
\eeq
where $i, \ j, \ \ell$ label vertices (sites) in $G$, $\sigma_i=1,...,q$ are
classical spin variables on these vertices, $\beta = (k_BT)^{-1}$, and $\langle
i j \rangle$ denote pairs of adjacent vertices. Without loss of generality, we
have taken the magnetic field $H$ to single out the spin value
$\sigma_i=1$. Let us introduce the notation
\beq
K = \beta J \ , \quad h = \beta H \ , \quad y = e^K \ , \quad v = y-1 \ , 
\quad w=e^h \ . 
\label{kdef}
\eeq
Thus, the physical ranges of $v$ are $v \ge 0$ for the Potts ferromagnet, and
$-1 \le v \le 0$ for Potts antiferromagnet. The weighted chromatic polynomial
is then obtained by choosing the antiferromagnetic sign of the spin-spin
coupling, $J < 0$ and taking $K \to -\infty$ while keeping $h=\beta H$
fixed. Since $K=\beta J$, the limit $K \to -\infty$ results if one takes $J \to
-\infty$ while holding $T$ and $H$ fixed and finite.  Alternatively, the limit
$K \to -\infty$ can be obtained by taking the zero-temperature limit $T \to 0$,
i.e., $\beta \to \infty$, with $J$ fixed and finite and $H \to 0$ so as to keep
$h = \beta H$ fixed and finite.  The limit $K \to -\infty$ guarantees that no
two adjacent spins have the same value, or, in the coloring context, no two
vertices have the same color.  One sees that in this statistical mechanics
context, it is the external magnetic field that produces the weighting that
favors or disfavors a given value for the spins $\sigma_i$.  Positive $H$ gives
a weighting that favors spin configurations in which spins have a particular
value, say 1, or equivalently, vertex colorings with this value of the color
assignment, while negative $H$ disfavors such configurations. For positive and
negative $H$, the corresponding ranges of $w$ are $w > 1$ and $0 \le w < 1$,
respectively.

The partition function $Z$ can be written in a manner that does not make
explicit reference to the spins $\sigma_i$ or the summation over spin
configurations, but instead as a sum of terms arising from spanning subgraphs
$G' \in G$.  The formula, obtained by F. Y. Wu, is \cite{wu78}
\beq
Z(G,q,v,w) = \sum_{G' \subseteq G} v^{e(G')} \
\prod_{i=1}^{k(G')} \Big ( q-1 + w^{n(G'_i)} \Big ) \ . 
\label{clusterw}
\eeq
This can be understood by writing Eqs. (\ref{z}) with (\ref{ham}) as 
\beq
Z=\sum_{ \{ \sigma_n \} }\Big [\prod_{\langle ij \rangle} (1+v\delta_{\sigma_i
  \sigma_j}) \Big ] \Big [ \prod_\ell e^{h\delta_{\sigma_\ell,1}} \Big ] \ . 
\label{vexp}
\eeq
If $h=0$, then each edge of a particular $G'$ gives a contribution of $v$ and
represents a spin configuration in which the spins on the ends of this edge
have the same value.  The spins in each component of $G'$ are connected by
edges, so they all have the same value, and there are $q$ possibilities for
this value.  In this case, from Eq. (\ref{vexp}) one sees that the resultant
term in summand of (\ref{clusterw}) is simply $v^{e(G')}q^{k(G')}$.  If $h \ne
0$, all of the spins in each connected component $G'_i$ of $G'$ have either the
value $\sigma_i=1$ or all of these spins have one of the other $q-1$ values. If
they all have the value 1, then each vertex in this $G'_i$ gives a contribution
of $w$, so from $G'_i$ one gets the contribution $w^{n(G'_i)}$, while if they
all have one of the other $q-1$ values, the contribution is 1. In general,
therefore, the contribution of the component $G'_i$ in $G'$ is
$(q-1+w^{n(G'_i)})$. Taking account of all of the $k(G')$ components in each
$G'$ gives the factor $\prod_{i=1}^{k(G')} \Big ( q-1 + w^{n(G'_i)} \Big )$,
which is then summed over all $G' \subseteq G$.  The Wu formula
(\ref{clusterw}) is a generalization of the Fortuin-Kasteleyn formula for the
zero-field model \cite{fk}.  The original definition of the Potts model,
(\ref{z}) and (\ref{ham}), requires $q$ to be in the set of positive integers
${\mathbb N}_+$.  This restriction is removed by Eq. (\ref{clusterw}).
Eq. (\ref{clusterw}) shows that $Z$ is a polynomial in the variables $q$, $v$,
and $w$, hence our notation $Z(G,q,v,w)$.

In the special case of zero external magnetic field, $H=0$, whence $w=1$, one
has the reduction to the Fortuin-Kasteleyn cluster formula \cite{fk} 
\beq
Z(G,q,v,1) = \sum_{G' \subseteq G} v^{e(G')} \ q^{k(G')} \ . 
\label{cluster}
\eeq
This zero-field Potts model partition function is equivalent to the Tutte
polynomial $T(G,x,y)$, defined by 
\beq
T(G,x,y) = \sum_{G' \subseteq G} (x-1)^{k(G')-k(G)} \, (y-1)^{c(G')} \ , 
\label{t}
\eeq
where $c(G')$ is the number of linearly independent cycles in $G'$, satisfying
$c(G')=e(G')+k(G')-n(G')$, and 
\beq
x = 1 + \frac{q}{v} \ . 
\label{x}
\eeq
(We remark that $k(G')-k(G)$ and $c(G')$ are the rank and co-rank of $G'$.) 
The equivalence is given by the relation
\beq
Z(G,q,v) = (x-1)^{k(G)}(y-1)^{n(G)} \, T(G,x,y) \ . 
\label{zt}
\eeq
In Ref. \cite{hl} we defined a generalization of the Tutte polynomial, 
\beqs
U(G,x,y,w) & = & (x-1)^{-k(G)}(y-1)^{-n(G)}\sum_{G' \subseteq G} (y-1)^{e(G')}
\ \times \cr\cr
& \times & \prod_{i=1}^{k(G')} (xy-x-y+w^{n(G'_i)}) \ .
\label{u}
\eeqs
This function satisfies $U(G,x,y,w)=(x-1)^{-k(G)}(y-1)^{-n(G)}Z(G,q,v,w)$ and
reduces to the Tutte polynomial if $w=1$: $U(G,x,y,1)=T(G,x,y)$.

The $K \to -\infty$ limit that yields the weighted chromatic polynomial is
equivalent to $v=-1$, so
\beq
Ph(G,q,w) = Z(G,q,-1,w) \ . 
\label{phz}
\eeq
Hence, a constructive formula for $Ph(G,q,w)$ is 
\beq
Ph(G,q,w) = \sum_{G' \subseteq G} (-1)^{e(G')} \
\prod_{i=1}^{k(G')} \Big ( q-1 + w^{n(G'_i)} \Big ) \ . 
\label{phclusterw}
\eeq
For the special case $h=0$, i.e., $w=1$, one thus has the result of
Eq. (\ref{phw1}).  The limit $h \to -\infty$, i.e., $w \to 0$, effectively
removes one of the possible values of the dynamical variables $\sigma_i$, or
equivalently, in the spanning subgraph formula (\ref{clusterw}), one of the
values of $q$, so
\beq
Z(G,q,v,0) = Z(G,q-1,v,1) \ .
\label{zw0}
\eeq
The special case of this for $v=-1$ is Eq. (\ref{phw0}).
It follows that each of the zeros of $Ph(G,q,1) \equiv P(G,q)$ in the complex
$q$ plane shifts to the right by one unit as one replaces the value $w=1$ by
$w=0$.  In the limit as $n \to \infty$, the accumulation set of the zeros,
${\cal B}_q$ also is replaced by its identical image shifted to the right in
the $q$ plane as one replaces $w=1$ by $w=0$.

\subsection{Results for Graphs with Loops, Multiple Edges, and Multiple 
Components}

If $G$ has any loop, defined as an edge that connects a vertex to itself, then
a proper $q$-coloring is impossible.  This is because such a $q$-coloring
requires that any two adjacent vertices have different colors, but since the
vertices connected by an edge are adjacent, the presence of a loop in $G$ means
that a vertex is adjacent to itself.  Thus,
\beq
Ph(G,q,w) = 0 \ {\rm if} \ G \ {\rm contains \ a \ loop} \ . 
\label{phloop}
\eeq
Hence, to avoid having $Ph(G,q,v)$ vanish trivially, we shall restrict our
analysis in this paper to loopless graphs $G$.  Accordingly, in the text below,
where $G=(V,E)$ is characterized as having a non-empty edge set $E \ne
\emptyset$, it is understood that $E$ does not contain any loops.

Another basic property of a chromatic polynomial is that as long as two
vertices are joined by an edge, adding more edges connecting these vertices
does not change the chromatic polynomial. This is clear from the fact that the
chromatic polynomial counts the number of proper $q$-colorings of the vertices
of $G$, and the relevant condition - that two adjacent vertices must have
different colors - is the same regardless of whether one or more than one edges
join these vertices.  Let us define an operation of ``\underline{r}eduction of
multiple \underline{e}dge(s)'' in $G$, denoted $R_E(G)$, as follows: if two
vertices are joined by a multiple edge, then delete all but one of these edges,
and carry out this reduction on all edges, so that the resultant graph $R_E(G)$
has only single edges. Then if $G$ is a graph that contains one or more
multiple edges joining some set(s) of vertices,
\beq
P(G,q) = P(R_E(G),q) \ . 
\label{pgme}
\eeq
Since the same proper $q$-coloring condition holds for the weighted chromatic
polynomial, we have
\beq
Ph(G,q,w) = Ph(R_E(G),q,w) \ . 
\label{phgme}
\eeq
Moreover, if $G$ consists of two disjoint parts, $G_1$ and $G_2$, then
$Ph(G,q,w)$ is simply the product $Ph(G,q,w)=Ph(G_1,q,w)Ph(G_2,q,w)$. Hence,
without loss of generality, we will generally restrict to connected graphs $G$.

\subsection{General Structural Properties of $Ph(G,q,w)$} 

Here we prove some general structural properties of $Ph(G,q,w)$ that hold for
an abitrary graph $G$. As discussed above, to avoid having $Ph(G,q,w)$ vanish
trivially, we take $G$ to be loopless, and without loss of generality, we
assume that $G$ is connected.  We first apply the proper $q$-coloring condition
to analyze properties of $Ph(G,q,w)$ for $q \in {\mathbb N}_+$.  Since this
proper $q$-coloring condition cannot be met for integer $q=1,...,\chi(G)-1$, it
follows that
\beq
Ph(G,q,w) \ {\rm contains \ a \ factor} \ \prod_{j=1}^{\chi(G)-1} (q-j) \ . 
\label{phqfactor}
\eeq
Provided that $G=(V,E)$ has at least one edge, i.e., $E \ne \emptyset$, the
proper $q$-coloring condition cannot be satisfied if $q=1$.  Hence, a corollary
of Eq. (\ref{phqfactor}) is
\beq
{\rm If} \ E \ne \emptyset, \ {\rm then} \ Ph(G,q,w) \ {\rm contains \ a \ 
factor} \ (q-1) \ . 
\label{qminus1factor}
\eeq
We can show that if $w=0$, then the factor $(q-1)$ is present even if $G$ does
not contain any edge. Using our previous result that $Z(G,q,v,0)=Z(G,q-1,v,1)$
and the fact that $Z(G,q,v,1)$ has a factor of $q$, we obtain the result that
$Z(G,q,v,0)$ contains the factor $q-1$ and hence $Ph(G,q,0)$ contains a factor
$(q-1)$.  More generally, since $Ph(G,q,0)=P(G,q-1)$ and $P(G,q-1)$ vanishes
for integer $q=1,...,\chi(G)$, it follows that
\beq
Ph(G,q,0) \ {\rm contains \ a \ factor} \ \prod_{j=1}^{\chi(G)}(q-j) \ . 
\label{phw0factors}
\eeq
Substituting $q=0$ in (\ref{clusterw}) and using the factorization
\beq
w^{n(G'_i)}-1 = (w-1)\sum_{\ell=0}^{n(G'_i)-1} w^\ell 
\label{wfac}
\eeq
proves that
\beq
Z(G,0,v,w) \ {\rm contains \ a \ factor \ of} \ (w-1) \ . 
\label{zq0}
\eeq
Setting $v=-1$, we thus deduce that \cite{hl} 
\beq
Ph(G,0,w) \ {\rm contains \ a \ factor \ of} \ (w-1) \ . 
\label{phq0}
\eeq
It is convenient to define the notation 
\beq
\tilde q = q-1 \ , \quad\quad \tilde w = w-1 \ . 
\label{tilde}
\eeq
From Eq. (\ref{clusterw}), it follows that we can write $Z(G,q,v,w)$ in several
equivalent ways:
\beqs
& & Z(G,q,v,w) = \sum_{r,t=0}^{n(G)} \, \sum_{s=0}^{e(G)} \ a_{r,s,t} \,
q^r v^s w^t  \ = \
\sum_{r,t=0}^{n(G)} \, \sum_{s=0}^{e(G)} \ b_{r,s,t} \, q^r y^s w^t  \cr\cr
& = &
\sum_{r,t=0}^{n(G)} \, \sum_{s=0}^{e(G)} \ c_{r,s,t} \,
\tilde q^{\, r} v^s w^t \ = \
\sum_{r,t=0}^{n(G)} \, \sum_{s=0}^{e(G)} \ d_{r,s,t} \, q^r v^s \tilde w^{\, t}
 \ ,
\label{zgenformwv}
\eeqs
where $a_{r,s,t}$, $b_{r,s,t}$, $c_{r,s,t}$, and $d_{r,s,t}$ are integers.
Some $a_{r,s,t}$ and $b_{r,s,t}$ can be negative, but, as we showed in
\cite{hl}, the nonzero $c_{r,s,t}$ and $d_{r,s,t}$ are positive. From these
equations, one infers corresponding ones for $Ph(G,q,w)$ by setting $v=-1$,
i.e., $y=0$. Note that in the polynomial $Z(G,q,v,w) = \sum_{r,t=0}^{n(G)} \, 
\sum_{s=0}^{e(G)} \ b_{rst} \, q^r y^s w^t$, clearly only the terms with 
$s=0$ contribute to $Ph(G,q,w)$.

From Eq. (\ref{clusterw}), it is evident that the term in $Ph(G,q,w)$ of
maximal degree in $q$, or equivalently, in $\tilde q$, arises from the
contribution of the spanning subgraph $G'$ with no edges, for which 
$k(G')=n(G)$.  This term is (with $n \equiv n(G)$)
\beq
(\tilde q + w)^n \ . 
\label{maxdegterm}
\eeq
It follows that 
\beq
a_{n,0,0} = b_{n,0,0}=c_{n,0,0}=d_{n,0,0}=1 \ . 
\label{an00}
\eeq
There are $e(G)$ spanning subgraphs $G'$ with one edge, since there are $e(G)$
ways of choosing this edge. Hence (with our restriction to loopless $G$), the
contribution of these $G'$ in (\ref{clusterw}) is
\beq
e(G)v(\tilde q + w^2)(\tilde q + w)^{n-2} \ . 
\label{oneedge}
\eeq
Expanding the terms in Eqs. (\ref{maxdegterm}) and (\ref{oneedge}) in powers of
$\tilde q$ and $w$, we find that the term in $Z(G,q,v,w)$ of degree $n-1$ in
$\tilde q$ is 
\beq
\Big ( e(G)v+nw \Big ) \, \tilde q^{\, n-1} \ . 
\label{qtildesecondhighest}
\eeq
Similarly, expanding the terms in Eqs. (\ref{maxdegterm}) and (\ref{oneedge}) 
in powers of $q$ and $w$, we find that the term in $Z(G,q,v,w)$ of degree 
$n-1$ in $q$ is 
\beq
\Big ( e(G)v+n(w-1) \Big ) \, q^{n-1} \ . 
\label{qsecondhighest}
\eeq

For our analysis below and for comparisons with chromatic polynomials, it will
be convenient to write $Ph(G,q,w)$ as a polynomial in $q$ with $w$-dependent
coefficients, which we denote $\alpha_{G,\ell}(w)$: 
\beq
Ph(G,q,w) = \sum_{j=0}^n \alpha_{G,n-j}(w) q^{n-j} \ . 
\label{phsum}
\eeq
From our discussion above, we have
\beq
\alpha_{G,n}=1 \ , 
\label{alphaqn}
\eeq
and, using also Eq. (\ref{phgme}), 
\beq
\alpha_{G,n-1} = - \Big ( e(R_E(G))+n(1-w) \Big ) \ . 
\label{alphaqnminus1}
\eeq
Moreover, from Eq. (\ref{phq0}), it follows that the $q^0$ term in
$Ph(G,q,w)$ contains a factor of $(w-1)$, i.e.,
\beq
\alpha_{G,0} \ {\rm contains \ a \ factor \ of} \ (w-1) \ . 
\label{alphaq0}
\eeq

It is also useful to express $Ph(G,q,w)$ as a polynomial in $w$ with
$q$-dependent coefficients, which we denote $\beta_{G,\ell}(q)$ (there should
not be confusion with $\beta=1/(k_BT)$):
\beq
Ph(G,q,w) = \sum_{j=0}^{d_w(G)} \beta_{G,j}(q) w^j \ , 
\label{phsumw}
\eeq
where $d_w(G) \equiv {\rm deg}_w(Ph(G,q,w))$ is the (maximal) degree of
$Ph(G,q,w)$ in $w$.  This degree, $d_w(G)$ is a $G$-dependent number less than
$n$.  To understand this, we recall that the maximum degree of $Z(G,q,v,w)$ in
$w$ is $n$.  This term is $y^{e(G)}w^{n} = (v+1)^{e(G)}w^{n}$ and corresponds
to all of the vertices having the same color, 1. However, the possibility that
all of the vertices have the same color, and, indeed, the possibility that any
adjacent vertices have the same color, are excluded for $Ph(G,q,w)$, as is
evident from the fact that the coefficient of $w^n$ vanishes for $v=-1$.
Hence, $d_w(G) < n$. We shall give this degree below for various families of
graphs.  Since all of the nontrivial graphs $G=(V,E)$ that we shall consider
have at least one edge, i.e., $E \ne \emptyset$, Eq. (\ref{qminus1factor})
shows that for these, $Ph(G,q,w)$ has the factor $(q-1)$.  In analyzing zeros
of $Ph(G,q,w)$ it will be convenient to separate this factor out, and we thus
define, for graphs containing at least one edge,
\beq
\beta_{G,j}(q) = (q-1)\bar \beta_{G,j}(q) \ , 
\label{betabar}
\eeq
so that
\beq
{\rm If} \ E \ne \emptyset, \ {\rm then} \quad 
Ph(G,q,w) = (q-1)\sum_{j=0}^{d_w(G)} \bar \beta_{G,j}(q) w^j\ . 
\label{phsumwbar}
\eeq
where $\bar\beta_{G,j}(q)$ are polynomials in $q$.  From Eq. (\ref{phsumw}) and
Eq. (\ref{phw0}), we obtain the relation
\beq
Ph(G,q,0)=\beta_{G,0}(q)=P(G,q-1) \ . 
\label{auxeq1}
\eeq
Combining this with Eq. (\ref{phw0factors}), we have the result that 
\beq
\beta_{G,0} \ {\rm contains \ a \ factor} \quad \prod_{j=1}^{\chi(G)}(q-j) \ . 
\label{beta0factors}
\eeq
Now, $Ph(G,q,1)=\sum_{j=1}^{d_w(G)} \beta_{G,j}$, but also 
$Ph(G,q,1)=P(G,q)$, so, using the fact that 
$P(G,q)=0$ for integer $q=0,...,\chi(G)-1$, we derive the following
factorization property for the sum of the $\beta_{G,j}$ coefficients: 
\beq
\sum_{j=1}^{d_w(G)} \beta_{G,j} \quad {\rm contains \ a \ factor} \quad 
\prod_{j=0}^{\chi(G)-1} (q-j) \ . 
\label{betasumfactors}
\eeq

\subsection{Absence of Deletion-Contraction Relation} 

For a graph $G$, we denote the graph obtained by deleting an edge $e \in E$ as
$G-e$ and the graph obtained by identifying the two vertices connected by this
edge $e$ as $G/e$.  The chromatic polynomial satisfies the deletion-contraction
relation $P(G,q,v) = P(G-e,q)-P(G/e,q)$. In contrast, for $w$ not equal to 1 or
0, the polynomial $Ph(G,q,w)$ does not, in general, satisfy this
deletion-contraction relation. It is of interest to examine the
quantities that measure the deviation from such a relation, namely 
\beq
\Delta Ph(G,e,q,w) = Ph(G,q,w)-\Big [ Ph(G-e,q,w) - Ph(G/e,q,w) \Big ] \ . 
\label{phdelcondif}
\eeq
We know that $\Delta Ph(G,e,q,w)$ contains a factor $w(w-1)$ since for $w=1$
and $w=0$, $Ph(G,q,w)$ is equal, respectively, to $P(G,q)$ and $P(G,q-1)$,
both of which do satisfy the deletion-contraction relation.  Furthermore,
because of Eq. (\ref{qminus1factor}), if $G$, $G-e$, and $G/e$ contain at least
one edge, then $\Delta Ph(G,e,q,w)$ contains the factor $(q-1)$.

As an illustration, using our explicit calculations given below for $n$-vertex
line graphs $L_n$ and circuit graphs $C_n$, we find the following results. For
the first two graphs, $L_3$ and $C_3$, the deletion and contraction on any 
edge gives the same result, so we need not specify which edge is involved.  We
find, for any edge $e$, 
\beq
\Delta Ph(L_3,e,q,w) = \Delta Ph(C_3,e,q,w) = -w(w-1)(q-1) 
\label{phline3_delcondif}
\eeq
and
\beq
\Delta Ph(C_4,e,q,w) = -w(w-1)(q-1)(q-2) \ . 
\label{phcyc4_delcondif}
\eeq
For $L_4$, denoting $e_{outer}$ as either of the two outer edges and $e_{mid}$
as the middle edge, we find 
\beq
\Delta Ph(L_4,e_{mid},q,w) = -w(w-1)(q-1)^2
\label{phline4_delcondif_emiddle}
\eeq
and
\beq
\Delta Ph(L_4,e_{outer},q,w) = -w(w-1)(q-1)(q+w-2) \ . 
\label{phline4_delcondif_eouter}
\eeq
It is straightforward to calculate similar differences $\Delta Ph(G,e,q,w)$ for
graphs with more vertices and edges, but these are sufficient to illustrate the
absence of a usual deletion-contraction relation for the weighted chromatic
polynomial.

\subsection{$T$, $P$, $U$, and $Ph$ Equivalence Classes} 

An important property of the weighted chromatic polynomial $Ph(G,q,w)$ is the
fact that it can distinguish between certain graphs that yield the same
chromatic polynomial $P(G,q)$.  More generally, an important property of the
partition function of the Potts model in a nonzero external magnetic field,
$Z(G,q,v,w)$, or equivalently, the function $U(G,x,y,w)$ that we defined in
Ref. \cite{hl}, is that $Z(G,q,v,w)$ and $U(G,x,y,w)$ distinguish between
graphs that yield the same zero-field Potts model partition function,
$Z(G,q,v,1)$ or equivalently, Tutte polynomial $T(G,x,y)$.  We begin with some
definitions.  Two graphs $G$ and $H$ are defined as (i) Tutte-equivalent
($T$-equivalent) if they have the same Tutte polynomial, or equivalently, the
same zero-field Potts model partition function, $Z(G,q,v,1)$; (ii)
$U$-equivalent if they have the same $Z(G,q,v,w)$; (iii) chromatically
equivalent ($P$-equivalent) if they have the same chromatic polynomial, 
$P(G,q)$, and (iv) $Ph$-equivalent if they have the same weighted chromatic
polynomial, $Ph(G,q,w)$.  

Let us give some examples.  Recall the definition that a tree graph is a
connected graph that contains no circuits (cycles).  The set of tree graphs
$\{G_{tree,n}\}$ with a given number, $n$, of vertices, forms a Tutte
equivalence class, with $T(G_{tree,n},x,y)=x^{n-1}$, or equivalently,
$Z(G,q,v,1)=q(q+v)^{n-1}$. However, the Potts partition function in a field,
$Z(G,q,v,w)$, or equivalently, the function $U(G,x,y,w)$ is able to distinguish
between different tree graphs in a Tutte-equivalence class. For instance,
consider the $n=4$ line graph $L_4$ and star graph $S_4$ (the graph with one
central vertex connected to three outer vertices by corresponding edges).
These have the same Tutte polynomial $T(L_4,x,y)=T(S_4,x,y)=x^3$, or
equivalently, the same zero-field Potts partition function
$Z(L_4,q,v,1)=Z(S_4,q,v,1)=q(q+v)^3$, but the full Potts partition functions,
$Z(L_4,q,v,w)$ and $Z(S_4,q,v,w)$ are different (see Eqs. (\ref{zline}) and
(\ref{zstar}) below).  Similarly, $L_4$ and $S_4$ are chromatically equivalent,
with $P(L_4,q)=P(S_4,q)=q(q-1)^3$ as a special case of the result
$P(G_{tree,n},q)=q(q-1)^{n-1}$ for any tree graph with $n$ vertices,
$G_{tree,n}$. However, from our calculations given below in (\ref{phline4}) and
(\ref{phstar4}), we find that $Ph(L_4,q,w)$ and $Ph(S_4,q,w)$ are different.
We reach the same conclusion for all of the tree graphs that we have studied,
i.e., although the set of tree graphs with a given number, $n$, of vertices,
forms a chromatic equivalence class, these graphs have different weighted
chromatic polynomials.  We will illustrate this below for $n=5$ and $n=6$.

A second set of examples involves graphs with multiple edges. Let us assume
that $G$ contains one or more multiple edges joining pair(s) of vertices. Such
graphs are not Tutte-equivalent, but, as noted above, are chromatically
equivalent.  Because the same proper $q$-coloring condition also holds for
weighted chromatic polynomials, these graphs are also in the same
$Ph$-equivalence class, as was stated in Eq. (\ref{phgme}).  A simple example
is provided by the line and cirtuit graphs with $n=2$ vertices, $L_2$ and
$C_2$, the latter of which has a double edge connecting the two vertices. One
has 
\beq
Z(L_2,q,v,w)=(q-1+w)^2+v(q-1+w^2)
\label{zline2}
\eeq
and
\beq
Z(C_2,q,v,w)=(q-1+w)^2+v(v+2)(q-1+w^2) \ , 
\label{zc2}
\eeq
so that
\beq
Z(C_2,q,v,w)-Z(L_2,q,v,w)=v(v+1)(q-1+w^2) \ . 
\label{zc2_minus_zline2}
\eeq
The fact that the difference in Eq. (\ref{zc2_minus_zline2}) vanishes for
$v=-1$, i.e., that $Ph(L_2,q,w)=Ph(C_2,q,w)$, is a special case of the general
result (\ref{phgme}).  

Because of the above-mentioned result that all $n$-vertex tree graphs are
chromatically equivalent, in conjunction with the property that $Ph(G,q,w)$ is
a chromatic polynomial for $w=1$ and $w=0$, it follows that the difference
between $Ph(G,q,w)$ and $Ph(H,q,w)$ between two chromatically equivalent graphs
$G$ and $H$ must vanish if $w=1$ or $w=0$. Since these are all polynomials, it
thus follows that the difference $Ph(G,q,w)-Ph(H,q,w)$ must have $w$ and $w-1$
as factors. Furthermore, if $q=1$, then
\beq
Z(G,1,v,w)=y^{e(G)}w^{n(G)} \ . 
\label{zq1}
\eeq
If $G$ has at least one edge, then $Z(G,1,v,w)=0$ if $y=0$, i.e., $v=-1$, 
Now in order to be chromatically equivalent, a necessary condition is that two
graphs $G$ and $H$ must have the same number of vertices, $n(G)=n(H)$, 
since the degree in $q$ of $P(G,q)$ is $n(G)$. An elementary property of
the chromatic polynomial $P(G,q)$, proved by iterative application of the
deletion-contraction theorem, is that the coefficient of the $q^{n(G)-1}$ 
term is $-e(R_E(G))$.  Therefore, another necessary condition that two graphs 
$G$ and $H$ be chromatically equivalent is that $e(R_E(G))=e(R_E(H))$. 
Now recall Eq. (\ref{qminus1factor}), according to which if $G$ contains at
least one edge, then $Ph(G,1,w)=0$.  Hence, if $G$ and $H$ are chromatically
equivalent, then either (i) neither contains any edges, in which case 
$Ph(G,q,w)=Ph(H,q,w)=(q-1+w)^n$, where $n=n(G)=n(H)$, or (ii) if $G$, and hence
$H$, contains at least one edge, $Ph(G,1,w)=Ph(H,1,w)=0$.  Hence, if $G$ and
$H$ are chromatically equivalent and contain at least one edge, then the
difference $Ph(G,q,w)-Ph(H,q,w)$ contains the factor $(q-1)$.  These results on
the factors of $Ph(G,q,w)-Ph(H,q,w)$ for chromatically equivalent graphs will
be evident in our explicit calculations to be presented below.

\subsection{On the Weighted Face Coloring Problem for Planar Graphs}

Let us consider a planar graph $G=(V,E)$.  We recall that the dual of this
graph, $G^*$, is the graph obtained from $G$ by associating a vertex of $G^*$
with each face of $G$ and connecting these vertices of $G^*$ with edges that
cross each edge of $G$.  There is thus a 1-1 isomorphism between the vertices,
edges, and faces of $G$ and the faces, edges, and vertices of $G^*$,
respectively.  A proper $q$ coloring of the faces of $G^*$ is a coloring of
these faces with $q$ colors subject to the constraint that no two faces that
are adjacent across the same edge have the same color. The (usual, unweighted)
chromatic polynomial $P(G,q)$ satisfies a duality property, namely that
$P(G,q)$ counts not just the proper $q$ colorings of the vertices of $G$, but
also, and equivalently, the proper $q$ colorings of the faces of $G^*$.  By the
same duality property, for a planar graph $G$, our weighted chromatic
polynomial $Ph(G,q,w)$ describes not just the weighted proper $q$ colorings of
the vertices of $G$ but also, and equivalently, the weighted proper $q$
colorings of the faces of $G^*$.

\section{Calculations of $Ph(G,q,w)$ for Some Families of Graphs}

In this section we give some illustrative explicit calculations of $Ph(G,q,w)$
for various families of graphs.  Although we generally consider connected
graphs, we note that for the graph $N_n$ consisting of $n$ vertices with no
edges,
\beq
Z(N_n,q,v,w)=Ph(N_n,q,w)=(q-1+w)^n \ . 
\label{phnn}
\eeq
We recall that a tree graph is defined as a connected graph with no
circuits. In the following text and in Appendix B we present results for 
the weighted chromatic polynomials of $n$-vertex tree graphs with $n$ up to 6.

\subsection{Line Graph $L_n$}

The line graph $L_n$ is the graph consisting of $n$ vertices with each vertex
connected to the next one by one edge.  One may picture this graph as forming a
line, whence the name.  For $n \ge 2$, the chromatic number is
$\chi(L_n)=2$. In \cite{zth} we gave a general formula for $Z(L_n,q,v,w)$, and
the special case $v=-1$ determines $Ph(L_n,q,w)$.  Let us define
\beq
T_{Z,1,0}= \left( \begin{array}{cc}
    q+v-1 & \ w \\
    q-1   & \ w(v+1) \end{array} \right )
\label{TTsq10}
\eeq
\beq
H_{1,0}= \left( \begin{array}{cc}
    1 & 0 \\
    0 & w \end{array} \right )
\label{H10}
\eeq
\beq
u_1 = {q-1 \choose 1}
\label{u1}
\eeq
and
\beq
s_1 = {1 \choose 1} \ . 
\label{s1}
\eeq
Then 
\beq
Z(L_n,q,v,w)= u_1^T \, H_{1,0} \, (T_{Z,1,0})^{n-1} \, s_1
\label{zline}
\eeq
and $Ph(L_n,q,w)=Z(L_n,q,-1,w)$. Since $e(L_n)=n-1$, we can apply
Eq. (\ref{alphaqnminus1}) to deduce that
\beq
\alpha_{L_n,n-1}=1+n(w-2) \ . 
\label{alpha_line_nminus1}
\eeq

From our general formula (\ref{zline}), evaluated at $v=-1$ to obtain
$Ph(L_n,q,w)$, we can derive some other corollaries concerning coefficients of
$Ph(L_n,q,w)$.  The maximal degree of $Ph(L_n,q,w)$ in $w$ is
\beq
{\rm deg}_w(Ph(L_n,q,w)) = \Big [\frac{n+1}{2} \Big ] \ , 
\label{maxdegwline}
\eeq
where here $[\nu]$ denotes the largest integer less than or equal to $\nu \in
{\mathbb R}$.  This contrasts with the fact that the highest power of $w$ in
$Z(L_n,q,v,w)$ for $v \ne -1$ is $n$.  The reason for this is that spin
configurations that would yield terms of degrees less than or equal to $n$ and
greater than the maximum in Eq. (\ref{maxdegwline}) are forbidden by the proper
$q$-coloring constraint.  If $n$ is odd, say $n=2m+1$, the coefficient of the
term in $Ph(L_{2m+1},q,w)$ of highest degree in $w$, namely the coefficient of
the term $w^{(n+1)/2}=w^{m+1}$, is
\beq
\beta_{L_{2m+1},m+1}=(q-1)^m \ . 
\label{betalnoddmax}
\eeq
If $n$ is even, say $n=2m$, the coefficient of the term in $Ph(L_{2m},q,w)$ of
highest degree in $w$, namely the coefficient of the term $w^{n/2}=w^m$, is
\beq
\beta_{L_{2m},m}=(q-1)^{m-1}\Big ( (m+1)q-2m \Big ) \ . 
\label{betalnevenmax}
\eeq
The coefficient of the $w^0$ term in $Ph(L_n,q,w)$ is 
\beq
\beta_{L_n,0}=(q-1)(q-2)^{n-1} \ . 
\label{betalnw0}
\eeq

We proceed to give some explicit results for $Ph(L_n,q,w)$ for various values
of $n$. The case $L_1=N_1$ is already covered by Eq. (\ref{phnn}).  For the
next few cases we list the explicit polynomials below, both in factored form
and in the form of Eq. (\ref{phsum}):
\beq
Ph(L_1,q,w) = q-1+w
\label{phline1}
\eeq
\beqs 
Ph(L_2,q,w) & = & (q-1)\Big [ q+2(w-1) \Big ] \cr\cr
            & = & q^2-(3-2w)q+2(1-w)
\label{phline2}
\eeqs
\beqs
Ph(L_3,q,w) & = & (q-1) \Big [ q^2+(3w-4)q + (w-1)(w-4) \Big ] \cr\cr
            & = & q^3-(5-3w)q^2+(w^2-8w+8)q-(w-1)(w-4) 
\label{phline3}
\eeqs
and
\beqs
Ph(L_4,q,w) & = & (q-1)(q+w-2) \Big [ q^2+(3w-4)q - 4(w-1) \Big ] \cr\cr
            & = & q^4-(7-4w)q^3+3(w^2-6w+6)q^2-(7w^2-26w+20)q \cr\cr
            & + & 4(w-1)(w-2) \ . 
\label{phline4}
\eeqs
Results for tree graphs with $n=5$ and $n=6$ vertices are given in 
Appendix B.

\subsection{Star Graphs $S_n$}

A star graph $S_n$ consists of one central vertex with degree $n-1$ connected
by edges with $n-1$ outer vertices, each of which has degree 1.  For $n \ge 2$,
the chromatic number is $\chi(S_n)=2$. We have derived the following general
formula for $Z(S_n,q,v,w)$:
\beq
Z(S_n,q,v,w) = \sum_{j=0}^{n-1} {n-1 \choose j} \, v^j \, (\tilde q + w^{j+1}) 
\, (\tilde q + w)^{n-1-j} \ , 
\label{zstar}
\eeq
where $\tilde q$ was given in Eq. (\ref{tilde}).  Evaluating this for $v=-1$
yields $Ph(S_n,q,w)$. The term in $Ph(S_n,q,w)$ of maximal degree in $w$
corresponds to a configuration in which all of the outer vertices are assigned
the color 1 and the central vertex of the star graph is assigned any of the
other $q-1$ colors.  For $n \ge 3$ where the star graphs are nondegenerate, 
this term is thus $(q-1)w^{n-1}$, so that, in particular, 
\beq
{\rm deg}_w(Ph(S_n,q,w)) = n-1 \ . 
\label{maxdegw_starn}
\eeq
(The graph $S_2$ is degenerate in the sense that it has no central vertex but
instead coincides with $L_2$.)  The graph $S_3$ is nondegenerate, and coincides
with $L_3$.  For $n=2$, the term in $Ph(S_2,q,w)$ of maximal degree in $w$,
namely the coefficient of the term $w$, is $2(q-1)$.  For $n \ge 3$, the
coefficient of the term in $Ph(S_n,q,w)$ of maximal degree in $w$, namely the
coefficient of the term $w^{n-1}$, is $(q-1)$. This is easily understood since
it corresponds to the assignment of the color 1 to each of the $n-1$ outer
vertices of the star graph $S_n$, which allows any of the remaining $(q-1)$
colors to be assigned to the central vertex of this graph. Because the number
of edges of the star graph is $e(S_n)=n-1$, it follows that
\beq
\alpha_{S_n,n-1}=1+n(w-2) \ . 
\label{alpha_star_nminus1}
\eeq
This coefficient is equal to $\alpha_{L_n,n-1}$. 

As an explicit example, for the graph $S_4$ we calculate
\beqs
Ph(S_4,q,w) & = & (q-1)\Big [ q^3 + 2(2w-3)q^2+(3w^2-14w+12)q+(w-1)(w^2-5w+8)
  \Big ] \cr\cr
            & = & q^4 - (7-4w)q^3 + 3(w^2-6w+6)q^2 - (-w^3+9w^2-27w+20)q \cr\cr
            & + & (1-w)(w^2-5w+8) \ . 
\label{phstar4}
\eeqs
Results for $S_n$ with $n=5$ and $n=6$ are given in Appendix B. 

\subsection{Distinguishing Between Some Chromatically Equivalent Graphs} 

Using the results given in the text and Appendix B for tree graphs with up to
six vertices, we now analyze the differences between the weighted chromatic
polynomials for tree graphs that are chromatically equivalent.  There are two
tree graphs with $n=4$ vertices, namely, $L_4$ and $S_4$.  In chemical
nomenclature, $L_4$ and $S_4$ correspond to the carbon atom backbones of the
alkanes n-butane and isobutane (i.e., 2-methylpropane). From
Eqs. (\ref{phline4}) and (\ref{phstar4}) we find
\beq
Ph(S_4,q,w)-Ph(L_4,q,w) = (q-1)w(w-1)^2 \ . 
\label{phstar4minusphline4}
\eeq
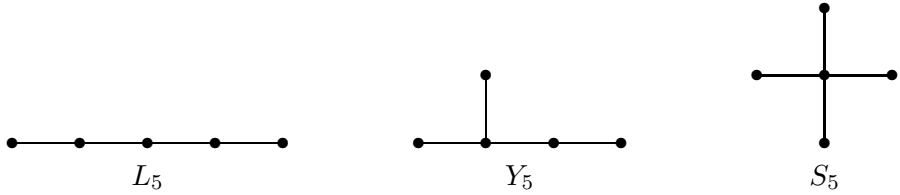
\begin{figure}[htbp]
\unitlength 0.9mm
\noindent
\begin{picture}(130,20)
\multiput(0,0)(10,0){5}{\circle*{1.6}}
\put(0,0){\line(1,0){40}}
\put(20,-5){\makebox(0,0){$L_5$}}
\multiput(60,0)(10,0){4}{\circle*{1.6}}
\put(70,10){\circle*{1.6}}
\put(60,0){\line(1,0){30}}
\put(70,0){\line(0,1){10}}
\put(75,-5){\makebox(0,0){$Y_5$}}
\multiput(110,10)(10,0){3}{\circle*{1.6}}
\multiput(120,0)(0,20){2}{\circle*{1.6}}
\put(110,10){\line(1,0){20}}
\put(120,0){\line(0,1){20}}
\put(120,-5){\makebox(0,0){$S_5$}}
\end{picture}
\vspace*{5mm}
\caption{\footnotesize{Tree graphs with $n=5$.}}
\label{tree_n5fig}
\end{figure}

There are three tree graphs with $n=5$ vertices, as shown in Fig.
\ref{tree_n5fig}, namely (i) the line graph $L_5$, (ii) a graph that we denote
$Y_5$, which is obtained by starting with the star $S_4$ and elongating one of
the edges by the addition of another vertex and edge, and (iii) the star graph
$S_5$.  We order this list in terms of graphs of increasing maximal vertex
degree $\Delta$; one has $\Delta=2, \ 3, \ 4$ for $L_5$, $Y_5$, and $S_5$,
respectively.  In chemical terminology, these correspond to the carbon atom
backbones of the alkanes (i) n-pentane, (ii) isopentane (i.e., 2-methylbutane),
and (iii) neopentane (i.e., 2,2-dimethylpropane), respectively.  From
Eqs. (\ref{phline5}), (\ref{phstar5}), and (\ref{phy5}), we calculate
\beq
Ph(S_5,q,w)-Ph(L_5,q,w) = (q-1)w(w-1)^2(3q+w-5)
\label{phstar5minusphline5}
\eeq
\beq
Ph(S_5,q,w)-Ph(Y_5,q,w) = (q-1)w(w-1)^2(2q+w-3)
\label{phstar5minusphy5}
\eeq
and thus
\beq
Ph(Y_5,q,w)-Ph(L_5,q,w) = (q-1)(q-2)w(w-1)^2 \ . 
\label{phy5minusline5}
\eeq
For these graphs, one observes that the differences between chromatically
equivalent graphs have a double zero at $w=1$. We find that this is also true
of the differences between weighted chromatic polynomials of tree graphs with
$n=6$ vertices, as discussed in Appendix B.

\subsection{Complete Graphs $K_n$}

The complete graph $K_n$ is the graph with $n$ vertices such that each vertex
is connected to every other vertex by one edge. The chromatic number is thus
$\chi(K_n)=n$. One has $e(K_n)={n \choose 2}$.  The simplest two cases coincide
with previously discussed graphs, namely the single vertex, $K_1=L_1$, for
which we gave $Ph(L_1,q,w)$ in Eq. (\ref{phline1}), and the
$n=2$ case, for which $K_2=L_2$ and $Ph(L_2,q,w)$ was given in
Eq. (\ref{phline2}). For general $n \ge 2$ we obtain the following
theorem:
\beq
Ph(K_n,q,w) = \Big [ \prod_{j=1}^{n-1}(q-j) \Big ](q+n(w-1)) \ . 
\label{phkn}
\eeq
Proof: To prove this, we begin by observing that because of the proper
$q$-coloring condition, $Ph(K_n,q,w)$ vanishes for all of the integer values
$q=1,...,n-1$ and hence must contain the factor $\prod_{j=1}^{n-1}(q-j)$.  The
proper $q$-coloring condition also means that only one vertex at most can be
assigned the color 1; hence the term in $Ph(K_n,q,w)$ of highest degree in the
variable $w$ has degree 1.  Since the maximal degree of $Ph(G,q,w)$ in the
variable $q$ is $n(G)$, it must be of the form
\beq
\Big [\prod_{j=1}^{n-1}(q-j) \Big ] (aq+bw+c) \ . 
\label{kngenaux}
\eeq
From Eq. (\ref{an00}), it follows that $a=1$. From Eq. (\ref{phw1}) we have
\beq
Ph(K_n,q,1)=P(K_n,q)=\prod_{j=0}^{n-1}(q-j) \ , 
\label{phknw1}
\eeq
which implies that $b=-c$, so the last factor in (\ref{kngenaux}) is 
$(q+b(w-1))$.  From Eq. (\ref{phw0}) we have 
\beq
Ph(K_n,q,0)=P(K_n,q-1)=\prod_{j=1}^{n}(q-j) \ , 
\label{phknw0}
\eeq
which implies that $b=n$, so that the additional factor is $(q+n(w-1))$. This
proves the result (\ref{phkn}). $\Box$  

A corollary of the theorem of Eq. (\ref{phkn}) is that
\beq
{\rm deg}_w(Ph(K_n,q,w))=1
\label{degwphkn}
\eeq
and, further, for $n \ge 2$, the term in $Ph(K_n,q,w)$ of maximal degree in $w$
has coefficient
\beq
\beta_{K_n,1}=n\prod_{j=1}^{n-1}(q-j) \ .
\label{phknwmaxbeta}
\eeq

\subsection{Wheel Graphs $Wh_n$}

The wheel graph $Wh_n$ is the graph obtained by joining one central vertex to
the $n-1$ vertices of the circuit graph $C_n$.  (This is the ``join'' of $K_1$
with $C_{n-1}$.) The central vertex can be regarded as forming the axle of the
wheel, while the $n-1$ vertices of the $C_{n-1}$ and their edges form the outer
rim of the wheel.  This is well-defined for $n \ge 3$, and in this range the
chromatic number is $\chi(Wh_n)=3$ if $n$ is odd and $\chi(Wh_n)=4$ if $n$ is
even.  The graph $Wh_3$ involves one double edge, while the $Wh_n$ graphs with
$n \ge 4$ have only single edges.  The first nondegenerate case is $Wh_4$,
which is the same graph as $K_4$.  We have given the general structure of
$Z(Wh_{n+1},q,v,w)$ in Ref. \cite{zth}, and this determines the structure of
$Ph(Wh_{n+1},q,w)$.  Reductions for $w=1$ and $w=0$ are given in Refs. 
\cite{dg,sdg}. For the nondegenerate cases $n \ge 3$, the number of edges
is $e(Wh_n)=2(n-1)$.  We calculate
\beqs
Ph(Wh_{n+1},q,w) & = & (q-1)\Big [ (\lambda_{Wh,+})^n + (\lambda_{Wh,-})^n \Big
] + (q-1)(q-3)(-1)^n \cr\cr
                 & + & w \Bigl [ (q-2)^n + (q-2)(-1)^n \Big ] \ , 
\label{phwheel}
\eeqs
where
\beq
\lambda_{Wh,\pm} = \frac{1}{2} \Big [ q-3 \pm \sqrt{A_{Wh}} \ \Big ]
\label{lamwheel}
\eeq
with
\beq
A_{Wh} = (q-3)^2+4w(q-2) \ . 
\label{awh}
\eeq
We note that $A_{Wh}$ is equal to $A_1$ (given in Eq. (\ref{a1})) with $q$
replaced by $q-1$, so that the eigenvalues $\lambda_{Wh,\pm}$ are the same as 
the eigenvalues $\lambda_{1,0,j}$, $j=1,2$ (given in Eq. (\ref{lam1d0j})) with
$q$ replaced by $q-1$:
\beq
\lambda_{Wh,\pm} = (\lambda_{1,0,j})_{q \to q-1} \ , 
\label{lamrel_wheelcircuit}
\eeq
where $\pm$ corresponds to $j=1,2$, respectively.  The $\lambda_{1,0,j}$,
$j=1,2$, enter in $Ph(L_n,q,w)$, given above, and $Ph(C_n,q,w)$, given in
Eq. (\ref{phcn}).  From these observations, it follows that
\beq
Ph(Wh_{n+1},q,w) = (q-1)Ph(C_n,q-1,w) + wP(C_n,q-1) \ . 
\label{phwhc}
\eeq
This relation makes the reductions of $Ph(Wh_{n+1},q,w)$ for $w=1$ and $w=0$
obvious; using $Ph(G,q,1)=P(G,q)$, one has 
\beqs
Ph(Wh_{n+1},q,1) & = & (q-1)P(C_n,q-1) + P(C_n,q-1) \cr\cr
                 & = & P(Wh_{n+1},q) = q \Big [ (q-2)^n + (q-2)(-1)^n \Big ]
\label{phwhw1}
\eeqs
and
\beqs
Ph(Wh_{n+1},q,0) & = & (q-1)P(C_n,q-2) \cr\cr
                 & = & P(Wh_{n+1},q-1) = 
(q-1) \Big [ (q-3)^n + (q-3)(-1)^n \Big ] \ . 
\label{phwhw0}
\eeqs
Further, from the values of $\chi(Wh_n)$ for odd and even $n$, it follows that 
\beq
{\rm If} \ n \ {\rm is \ odd} \ Ph(Wh_n,q,w) \ {\rm contains \ a \ factor}
\quad  (q-1)(q-2) \ . 
\label{pwheel_nodd_factors}
\eeq
and
\beq
{\rm If} \ n \ {\rm is \ even} \ Ph(Wh_n,q,w) \ {\rm contains \ a \ factor}
\quad  (q-1)(q-2)(q-3) \ .  
\label{pwheel_neven_factors}
\eeq
Although $Wh_3$ differs from $C_3=K_3$ in having one
double edge, Eq. (\ref{phgme}) shows that $Ph(Wh_3,q,w)=Ph(C_3,q,w)$, where
$Ph(C_3,q,w)$ was given above in Eq. (\ref{phc3}). Furthermore, the graph
$Wh_4$ is the same as $K_4$, so $Ph(Wh_4,q,w)=Ph(K_4,q,w)$, where $Ph(K_n,q,w)$
was given above in Eq. (\ref{phkn}). 

Since the number of edges in the wheel graph $e(Wh_{n+1})=2n$, we can apply 
Eq. (\ref{alphaqnminus1}) to deduce that 
\beq
\alpha_{Wh_{n+1},n}=-\Big ( 3n+1 -(n+1)w \Big ) \ . 
\label{alphanminus1_wheel}
\eeq

For the following we again assume that $n \ge 3$ so that the $Wh_n$ graph is 
well-defined.  The highest power of $w$ in $Ph(Wh_n,q,w)$ is
\beq
{\rm deg}_w(Ph(Wh_n,q,w)) = \Big [ \frac{n-1}{2} \Big ] \ . 
\label{wdegmax_wheel}
\eeq
If $n$ is even, say $n=2m$, then the coefficient of the term in $Ph(Wh_n,q,w)$
of maximal degree, namely the coefficient of the term $w^{m-1}$, is
\beq
\beta_{Wh_{2m},m-1}=(2m-1)(q-1)(q-2)^{m-1}(q-3) \ . 
\label{beta_wheelmaxneven}
\eeq
If $n$ is odd, say $n=2m+1$, then the coefficient of the term in $Ph(Wh_n,q,w)$
of maximal degree, namely the coefficient of the term $w^m$, is
\beq
\beta_{Wh_{2m+1},m}=2(q-1)(q-2)^m \ . 
\label{beta_wheelmaxnodd}
\eeq

As an illustration, we display $Ph(Wh_5,q,w)$ below:
\beqs
Ph(Wh_5,q,w) & = & (q-1)(q-2)
\Big [ q^3-5(2-w)q^2 + (2w^2-29w+34)q -(w-1)(4w-39) \Big ] \cr\cr
             & = & q^5 -(13-5w)q^4+2(w^2-22w+33)q^3-(10w^2-140w+161)q^2 \cr\cr
             & + & (16w^2-187w+185)q-2(w-1)(4w-39) \ . 
\label{phwheel5}
\eeqs

\section{$Ph(G,q,w)$ for Lattice Strip Graphs with Periodic Longitudinal
Boundary Conditions} 

\subsection{General Structure} 

In \cite{hl,zth} we have given a general structural formula for $Z(G_s, L_y
\times m,BC,q,v,w)$ on strip graphs $G_s$ of width $L_y$ vertices and
length $L_x$, with cyclic (cyc.) or M\"obius (Mb) boundary conditions (BC's). 
For cyclic strips the special case of this structural formula is
\beq
Ph(G_s, L_y \times m,cyc.,q,w) 
 = \sum_{d=0}^{L_y} \tilde c^{(d)} \sum_{j=1}^{n_{Ph}(L_y,d)}
[\lambda_{G_s,L_y,d,j}(q,w)]^m \ , 
\label{phsumcyc}
\eeq
where $m=L_x$ for strips of the square and triangular lattices and $m=L_x/2$
for strips of the honeycomb lattice. The coefficients $\tilde c^{(d)}$ are
given by
\beq
\tilde c^{(d)} = \sum_{j=0}^d (-1)^j {2d-j \choose j}(q-1)^{d-j} \ . 
\label{ctilde}
\eeq
The first few of these coefficients are $\tilde c^{(0)} = 1$, $\tilde c^{(1)} =
q-2$, $\tilde c^{(2)} = q^2-5q+5$, etc. For M\"obius strips, there is a
switching of certain $\tilde c^{(d)}$'s as specified in general in \cite{zth},
using the same methods that we employed in \cite{cf}, as specified by Eqs. 
(2.30)-(2.32) and the $v=-1$ special case of Eq. (2.33)) of \cite{zth}). 

The numbers $n_{Zh}(L_y,d)$ of $\lambda$'s corresponding to each $\tilde
c^{(d)}$ in the general Potts model partition function are reduced for the
special case $v=-1$ of interest here.  By coloring combinatoric arguments
similar to those used in \cite{cf} and \cite{zth} we determine the
$n_{Ph}(L_y,d)$ as follows.  The numbers $n_{Ph}(L_y,d)$ are identically zero
for $d > L_y$, and
\beq
n_{Ph}(L_y,L_y)=1
\label{nphlydly}
\eeq
\beq
n_{Ph}(L_y,L_y-1)=2L_y
\label{nphlydlym1}
\eeq
\beq
n_{Ph}(L_y+1,0)=n_{Ph}(L_y,0)+n_{Ph}(L_y,1)
\label{nphlyd0}
\eeq
and, for $1 \le d \le L_y+1$, 
\beq
n_{Ph}(L_y+1,d)=n_{Ph}(L_y+1,d-1)+2n_{Ph}(L_y,d)+n_{Ph}(L_y,d+1) \ . 
\label{nphlyd}
\eeq
The $n_{Ph}(L_y,d)$ satisfy the identity 
\beq
\sum_{d=0}^{L_y} \tilde c^{(d)} n_{Ph}(L_y,d) = P(T_{L_y},q) = q(q-1)^{L_y-1} \
. 
\label{eqcly}
\eeq
Indeed, one method of calculating the $n_{Ph}(L_y,d)$ is to differentiate this
equation $L_y$ times.  One thereby obtain $L_y+1$ linear equations in the
$L_y+1$ unknowns $n_{Ph}(L_y,d)$, $d=0,1,...,L_y$; solving these equations
yields the results given above.  We note that
\beq
n_{Ph}(L_y,0) = C_{L_y-1} + C_{L_y} \ , 
\label{nphlyd0aux}
\eeq
where $C_n$ is the Catalan number, 
\beq
C_n = \frac{1}{n+1}{2n \choose n} \ . 
\label{catalan}
\eeq
(No confusion should result with the use of $C_n$ to mean both Catalan number
and the cyclic graph with $n$ vertices, since the context makes clear which is
meant.)  We recall that the partition function of the zero-field Potts model on
cyclic strips of the square lattice (as well as other lattices) has the
structure \cite{saleur,cf}
\beq
Z(G_s, L_y \times m,cyc.,q,v) 
 = \sum_{d=0}^{L_y} c^{(d)} \sum_{j=1}^{n_Z(L_y,d)}
[\lambda_{Z,G_s,L_y,d,j}(q,v)]^m \ , 
\label{zgsumcyc}
\eeq
where the coefficients $c^{(d)}$ are given by Eq. (\ref{ctilde}) with $q \to
q+1$ and 
\beq
n_Z(L_y,d)=\frac{(2d+1)}{(L_y+d+1)}{2L_y \choose L_y-d}
\label{nzlyd}
\eeq
for $0 \le d \le L_y$.  For this $h=0$ case, the total number of distinct
eigenvalues that enter in Eq. (\ref{zgsumcyc}) for a lattice strip, i.e., 
\beq 
N_{Z,L_y} = \sum_{d=0}^{L_y} n_Z(L_y,d) \ , 
\label{nztoth0}
\eeq
is
\beq
N_{Z,L_y} = {2L_y \choose L_y} \quad {\rm for} \quad h=0  \ . 
\label{nztot}
\eeq

We find an interesting relation connecting the numbers
$n_{Ph}(L_y,d)$ with the corresponding numbers $n_Z(L_y,d)$ for the zero-field
Potts model, namely, for $L_y \ge 2$, 
\beq
n_{Ph}(L_y,d) = n_Z(L_y,d) + n_Z(L_y-1,d) \ . 
\label{nphzlyd_relation}
\eeq

From our determination of the $n_{Ph}(L_y,d)$, we next calculate the total
number 
\beq
N_{Ph,L_y} = \sum_{d=0}^{L_y} n_{Ph}(L_y,d) \ . 
\label{nphlytotsum}
\eeq
We find
\beq
N_{Ph,L_y} = {2L_y \choose L_y} + {2(L_y-1) \choose L_y-1} \ . 
\label{nphlytot}
\eeq
From the relation (\ref{nphzlyd_relation}) it follows that the total number
$N_{Ph,L_y}$ satisfies the relation, for $L_y \ge 2$, 
\beq
N_{Ph,L_y} = N_{Z,L_y} + N_{Z,L_y-1} \ , 
\label{nphztot_relation}
\eeq
as is evident in Eq. (\ref{nphlytot}).  We list the $n_{Ph}(L_y,d)$ and
$N_{Ph,L_y}$ for $1 \le L_y \le 8$ in Table \ref{nphtable}.  For purposes of
comparison, we include tables of $n_P(L_y,d)$, $N_{P,L_y}$, and $n_Z(L_y,d)$
for both $h=0$ and $h \ne 0$ (from \cite{zth}) in Appendix A.

\begin{table}
\caption{\footnotesize{Table of numbers $n_{Ph}(L_y,d)$ and their sums,
$N_{Ph,L_y}$ for strips of the lattice $\Lambda$ (square, triangular, or
 honeycomb).  Blank entries are zero.  See text for further discussion.}}
\begin{center}
\begin{tabular}{|c|c|c|c|c|c|c|c|c|c|c|}
\hline\hline
$L_y  \ \backslash \ d$
   & 0    & 1    & 2    & 3    & 4    & 5   & 6   & 7  & 8 & $N_{Ph,L_y}$ 
\\ \hline\hline
1  & 2    & 1    &      &      &      &     &     &    &   & 3 
\\ \hline
2  & 3    & 4    & 1    &      &      &     &     &    &   & 8
\\ \hline
3  & 7    & 12   & 6    & 1    &      &     &     &    &   & 26
\\ \hline
4  & 19   & 37   & 25   & 8    & 1    &     &     &    &   & 90
\\ \hline
5  & 56  & 118   & 95   & 42   & 10   & 1   &     &    &   & 322
\\ \hline
6  & 174 & 387   & 350  & 189  & 63   & 12  & 1   &    &   & 1176
\\ \hline
7  & 561 & 1298  & 1276 & 791  & 327  & 88  & 14  & 1  &   & 4356
\\ \hline
8  & 1859& 4433  & 4641 & 3185 & 1533 & 517 & 117 & 16 & \ 1 \ & 16302
\\ \hline\hline
\end{tabular}
\end{center}
\label{nphtable}
\end{table}

Let us denote $N_{Z,L_y}$ for $h \ne 0$ as $N_{Zh,L_y}$ for notational clarity,
to distinguish this from $N_Z(L_y,d)$ for $h=0$. For the Potts model in a
nonzero field, we have found \cite{hl,zth}
\beq
N_{Zh,L_y} = \sum_{j=0}^{L_y} {L_y \choose j} {2j \choose j} 
\quad {\rm for} \quad h \ne 0 \ . 
\eeq
Concerning the relative sizes of $N_{P,L_y}$, $N_{Ph,L_y}$,
$N_{Z,L_y}$ and $N_{Zh,L_y}$, we have, for $L_y=1$,
$N_{P,1}=N_{Z,1} = 2 < N_{Ph,1}=N_{Zh,1} = 3$ and the inequality
\beq
N_{P,L_y} < N_{Z,L_y} < N_{Ph,L_y} < N_{Zh,L_y} \quad {\rm for} \quad L_y \ge 2
\ . 
\label{ntotineq}
\eeq
For example, this set of four numbers is (4,6,8,11) and (10,20,26,45) for
$L_y=2$ and $L_y=3$, respectively. 

For large strip width $L_y$, $N_{Ph,L_y}$ has the same 
general asymptotic behavior as $N_{Z,L_y}$:
\beq
N_{Ph,L_y} \sim {\rm const.} \times {L_y}^{-1/2} \, 4^{L_y} \quad 
{\rm as} \ \ L_y \to \infty \ .
\label{nphtotasymp}
\eeq

\subsection{Circuit Graphs $C_n$}

The circuit graph $C_n$, or equivalently, the 1D lattice with periodic boundary
conditions, has chromatic number $\chi(C_n)=2$ if $n$ is even and $\chi(C_n)=3$
if $n \ge 3$ is odd.  (The case $n=1$ is a single vertex with a loop, for which
there is no proper $q$-coloring, so $Ph(C_1,q,w)$ vanishes identically.)  The
polynomial $Ph(C_n,q,w)$ for the circuit graph $C_n$, can be obtained from the
calculations of $Z(C_n,q,v,w)$ \cite{gu,zth} by setting $v=-1$.  Expressed in
our present notation, it is
\beq
Ph(C_n,q,w) = (\lambda_{1,0,1})^n + (\lambda_{1,0,2})^n + 
(q-2)(\lambda_{1,1})^n \ , 
\label{phcn}
\eeq
where
\beq
\lambda_{1,0,j} = \frac{1}{2}\Big [q-2 \pm \sqrt{A_1} \ \Big ]
\label{lam1d0j}
\eeq
where the $\pm$ sign corresponds to $j=1,2$, 
\beq
A_1 = (q-2)^2+4(q-1)w
\label{a1}
\eeq
and
\beq
\lambda_{1,1}=-1 \ . 
\label{lam11}
\eeq
From the values of $\chi(C_n)$ given above, it follows that, in addition to the
general factor of $(q-1)$ present for any $n$, if $n$ is odd, $Ph(C_n,q,w)$
contains a factor of $(q-2)$. Since $e(C_n)=n$, it follows that for $n \ge 3$
\beq
\alpha_{C_n,n-1}=n(w-2) \ . 
\label{alpha_cn_nminus1}
\eeq
(For $n=2$, $C_2$ has a double edge, so one uses Eq. (\ref{alphaqnminus1}) to
obtain $\alpha_{C_2,1}=-(3-2w)$.)  We exhibit $Ph(C_n,q,w)$ for $2 \le n
\le 5$ below:
\beq
Ph(C_2,q,w) = Ph(L_2,q,w) =  (q-1)[q+2(w-1)] = q^2 -(3-2w)q +2(1-w) 
\label{phc2}
\eeq
\beqs
Ph(C_3,q,w) & = & (q-1)(q-2) \Big [ q+3(w-1) \Big ] \cr\cr
            & = & q^3-3(2-w)q^2+(11-9w)q-6(1-w)
\label{phc3}
\eeqs
\beqs
Ph(C_4,q,w) &=& (q-1) \Big [ q^3+(4w-7)q^2+(2w^2-16w+17)q-2(w-1)(w-7)\Big ] 
\cr\cr
            & = & q^4-4(2-w)q^3+2(w^2-10w+12)q^2-(4w^2-32w+31)q \cr\cr
            & + & 2(w-1)(w-7)
\label{phc4}
\eeqs
and
\beqs
Ph(C_5,q,w) & = & (q-1)(q-2)\Big [ q^3+(5w-7)q^2+(5w^2-20w+17)q-
                  5(w-1)(w-3) \Big ] \cr\cr
            & = & q^5-5(2-w)q^4+5(w^2-7w+8)q^3-10(2w^2-9w+8)q^2 \cr\cr
            & + & (25w^2-100w+79)q-10(w-1)(w-3)  \ . 
\label{phc5}
\eeqs

In the special case of zero-field, $h=0$, i.e., $w=1$, $\lambda_{1,0,1}=q-1$
while $\lambda_{1,0,2}$ becomes equal to $\lambda_{1,1}$.  Thus, a
``transmigration'' process occurs in which one of the $\lambda$'s associated
with the coefficient $\tilde c^{(d)}$ of degree $d=0$ becomes equal to, and
hence can be grouped with, a $\lambda$ associated with a coefficient 
$\tilde c^{(d)}$ with a different degree $d$, here $d=1$. 
Hence, one has the reduction
\beq
Ph(C_n,q,1) = P(C_n,q) = (q-1)^n + (q-1)(-1)^n \ . 
\label{pcn}
\eeq
For $w=0$, $\lambda_{1,0,1}=q-2$ while $\lambda_{1,0,2}=0$, so that
$Ph(C_n,q,0)=Ph(C_n,q-1,1)$, in agreement with the general relation
(\ref{phw0}).

As an application of our result (\ref{qminus1factor}) above, it follows that 
$Ph(C_n,q,w)$ contains the factor $(q-1)$.  We note some additional 
factorization properties and special values of $Ph(C_n,q,w)$:
\beq
{\rm If} \ n \ {\rm is \ odd, \ then} \ Ph(C_n,q,w) \ {\rm contains \ the \ 
factor} \ (q-2) \ . 
\label{phcn_n_odd}
\eeq

For the $q=2$ case, 
\beq
Ph(C_n,2,w) = \Big [ 1 + (-1)^n \Big ] w^{n/2} \ . 
\label{phcnq2}
\eeq

The highest power of $w$ in $Ph(C_n,q,w)$ is 
\beq
{\rm deg}_w(Ph(C_n,q,w)) = \Big [ \frac{n}{2} \Big ] \ , 
\label{hipow_cn}
\eeq
where here $[\nu]$ denotes the integral part of $\nu$.  This contrasts with the
fact that the highest power of $w$ in $Z(C_n,q,v,w)$ for $v \ne -1$ is $n$.
The reason for this is that spin configurations that would yield terms
proportional to $w^p$ with $n/2 < p \le n$ for even $n$ and $(n-1)/2 < p \le n$
for odd $n$ are forbidden by the proper $q$-coloring constraint.  For $n$ even,
say $n=2m$, the term in $Ph(C_{2m},q,w)$ of maximal degree in $w$, namely 
$w^m$, has coefficient $2(q-1)^m$.  For $n$ odd, say $n=2m+1$ with $m \ge 1$,
the term in $Ph(C_{2m+1},q,w)$ of maximal degree in $w$, namely $w^m$, has
coefficient $(2m+1)(q-1)^m(q-2)$. 

If and only if $w=1$, then $Ph(C_n,q,1)=P(C_n,q)$ contains $q$ as a factor.
For this zero-field case $w=1$ we also recall that 
$Ph(C_n,q,1) = P(C_n,q)$ contains $q(q-1)$ as a factor and, furthermore, if $n
\ge 3$ is odd, then $Ph(C_n,q)$ also contains $(q-2)$ as a factor.

\subsection{$L_y=2$ Cyclic Strip}

We denote the cyclic and M\"obius strips of the square lattice of width 
$L_y=2$ and length $L_x=m$ as the ladder graph $L_m$ and the M\"obius ladder
graph $ML_m$.  For both of these our general structure determination above
gives $n_{Ph}(2,0)=3$, $n_{Ph}(2,1)=4$, and $n_{Ph}(2,2)=1$, for a total of 
$N_{Ph,2}=8$ terms.  The weighted coloring polynomial for $L_m$ is 
\beq
Ph(L_m,q,w) = \sum_{d=0}^2 \tilde c^{(d)} \sum_{j=1}^{n_{Ph}(2,d)} 
(\lambda_{2,d,j})^m \ , 
\label{ph2}
\eeq
where
\beq
\lambda_{2,0,1} = w(1-q)
\label{plam2d01}
\eeq
\beq
\lambda_{2,0,j} = \frac{1}{2}\Big [ q^2+(w-5)q+7-w \pm \sqrt{A_2} \ \Big ] \ , 
\quad j=2,3 
\label{plam2d023}
\eeq
\beqs
A_2 & = & q^4+6q^3w+q^2w^2-10q^3-36q^2w-2qw^2+39q^2 \cr\cr
    & + & 72qw+w^2-70q-50w+49
\label{a2}
\eeqs
\beq
\lambda_{2,1,j} = -\frac{1}{2}\Big [q-2 \pm \sqrt{A_3} \ \Big ] \ , \quad 
j=1,2
\label{plam2d112}
\eeq
\beq
A_3 = q^2+4(q-1)(w-1)
\label{a3}
\eeq
\beq
\lambda_{2,1,j} = -\frac{1}{2} \Big [ q-4 \pm \sqrt{A_4} \ \Big ] \ , \quad 
j=3,4
\label{plam2d123}
\eeq
\beq
A_4 = q^2+4q(w-2)+4(4-3w)
\label{a4}
\eeq
and
\beq
\lambda_{2,2}=1 \ . 
\label{plam22}
\eeq
In Eq. (\ref{plam2d123}), $j=3$ and $j=4$ apply for the $+$ and $-$ sign
choices, respectively, and similarly for the other equations. Results
for $Z(L_m,q,v,w)$ are given in Refs. \cite{mirza} and \cite{zth}. The weighted
chromatic polynomial for the $L_y=2$ M\"obius strip of the square lattice is
obtained by applying the results of Ref. \cite{zth}. 

For $w=1$, these $\lambda$'s reduce as follows: 
\beq
\lambda_{2,0,1} \to 1-q
\label{plam201w1}
\eeq
\beq
\lambda_{2,0,2} \to q^2-3q+3
\label{plam202w1}
\eeq
\beq
\lambda_{2,0,3} \to 3-q
\label{plam203w1}
\eeq
\beq
\lambda_{2,1,1} \to 1-q
\label{plam211w1}
\eeq
\beq
\lambda_{2,1,2} \to 1
\label{plam212w1}
\eeq
\beq
\lambda_{2,1,3} \to 3-q
\label{plam213w1}
\eeq
\beq
\lambda_{2,1,4} \to 1 \ . 
\label{plam214w1}
\eeq
Thus, a transmigration process of $\lambda$'s occurs here, just as it did for
the $L_y=1$ case; (i) one of the three $\lambda$'s in the $d=0$ subspace
reduces to the single $\lambda$, $q^2-3q+3$, in the $d=0$ subspace for the
chromatic polynomial $P(L_m,q)$, while the other two become equal to the two
$\lambda$'s in the $d=1$ subspace of $P(L_m,q)$; (ii) two of the four
$\lambda$'s in the $d=1$ subspace reduce to the two $\lambda$'s, $3-q$ and
$1-q$, in this subspace for $P(L_m,q)$, while the other two become equal to the
single $\lambda=1$, in the $d=2$ subspace for $P(L_m,q)$.  Hence, we have the
reduction
\beqs
Ph(L_m,q,1) & = & (1-q)^m + (q^2-3q+3)^m + (3-q)^m + 
\tilde c^{(1)} \Big [ (1-q)^m + 2 + (3-q)^m \Big ] + \tilde c^{(2)} \cr\cr
& = & (q^2-3q+3)^m + c^{(1)}\Big [ (3-q)^m + (1-q)^m \Big ] + c^{(2)} \ . 
\label{p2duced}
\eeqs

\section{Some Properties of the Zeros of $Ph(G,q,w)$} 

\subsection{Zeros of $Ph(G,q,w)$ in $q$ as Functions of $w$} 

Here we discuss the zeros of $Ph(G,q,w)$ in $q$ as a function of $w$ for some
illustrative graphs $G$.  Since the maximal degree of $Ph(G,q,w)$ in the
variable $q$ is $n(G)$, it has this number of zeros in the variable $q$.  In
contrast, as is evident from our explicit calculations above, the maximal
degree of $Ph(G,q,w)$ in the variable $w$ depends on details of $G$. As is
true for any polynomial, the positions of the zeros of $Ph(G,q,w)$ are
continuous functions of $q$ for fixed $w$ and continuous functions of $w$ for
fixed $q$.  As noted above, for any graph $G$ with at least one edge,
$Ph(G,q,w)$ contains the factor $(q-1)$, so it has a fixed zero at $q=1$. A
general statement is that since $Ph(G,q,1)=P(G,q)$ and $Ph(G,q,0)=P(G,q-1)$, it
follows that each zero of $Ph(G,q,w)$ shifts horizontally to the right by one
unit in the complex $q$ plane if one replaces $w=1$ by $w=0$.

One relevant quantity of interest is the maximal real zero of $Ph(G,q,w)$,
which we denote $q_{mrz}(G)$. This is related to chromatic number of the graph
$G$, $\chi(G)$, because, by the definition of $q_{mrz}(G)$, $Ph(G,q,w)$ is
nonzero for real $q > q_{mrz}(G)$, and it must be positive since for a given
$G$ and $w$, if $q$ is sufficiently large, then $Ph(G,q,w)$ is positive.
Hence, in addition to $\chi(G)$, which is fixed for a given $G$, the quantity
$q_{mrz}(G)$ serves as a $w$-dependent measure of the ability to perform a
proper vertex coloring of this graph.  As a corollary of the discussion above,
a general result is that $q_{mrz}(G)$ shifts one unit to the right as $w$
decreases from 1 to 0, as a consequence of (\ref{phw1}) and (\ref{phw0}).  

Let us consider some simple examples. We have
\beq
q_{mrz}(L_1) = 1-w \ . 
\label{qmrzline1}
\eeq
This increases from 0 to 1 as $w$ decreases from 1 to 0 in the DFCP interval
and decreases from 0 through negative values as $w$ increases above 1 in the
FCP interval. For $L_2$, 
\beq
q_{mrz}(L_2) = 2(1-w) \ . 
\label{qmrzline2}
\eeq
This increases from 0 to 2 as $w$ decreases from 1 to 0 in the DFCP interval
and decreases from 0 through negative values as $w$ increases from 1 in the
FCP interval. This example also illustrates how the multiplicity of zeros can
change as a function of $w$; for $w=1/2$, $Ph(L_2,q,1/2)$ has two coincident
zeros at $q=1$.

For $L_3$, the situation is more complicated.  The expression for $Ph(L_3,q,w)$
is given in Eq. (\ref{phline3}).  In addition to the fixed zero at $q=1$,
$Ph(L_3,q,w)$ has two other zeros, which occur at the values
\beq
q_{L3z,\pm} = \frac{1}{2}\Big [ 4-3w \pm \sqrt{w(5w-4)} \ \Big ] \ . 
\label{qmrzpm}
\eeq
For $w=1$, these reduce to $q=1$ and $q=0$ for the $\pm$ signs.  As $w$
decreases from 1, the zero at $q=0$ increases while the zero at $q=1$
decreases.  As $w$ decreases through the value $w=4/5$, these two zeros meet at
$q=4/5$, and then move off the real axis as a complex-conjugate pair as $w$
decreases further in the interval $0 < w < 4/5$.  The magnitudes of the
imaginary parts of these complex zeros increase to maximal values as $w$
decreases through the value $w=2/5$ and then decrease toward zero. As $w$
decreases through the value $w=0$, these zeros return to the real axis,
becoming a double zero at $q=2$.  Thus, for $w=1$, $q_{L3z,+}=1$, while
for $w=0$, $q_{L3z,+}=2$.  However, $q_{L3z,+}$ does not increase monotonically
from 1 to 2 as $w$ decreases from 1 to 0; instead, it actually decreases from 1
to 4/5 as $w$ decreases from 1 to 4/5, while $q_{L3z,-}$ increases from 0 to
4/5.  As $w$ decreases below 4/5, $q_{L3z,\pm}$ form a pair of
complex-conjugate roots, as noted. Hence, for $w$ in the DFCP interval, the
fixed zero at $q=1$ is $q_{mrz}(L_3)$.  As $w$ passes through the value $w=0$,
$q_{rmz}(L_3)$ jumps discontinuously from the fixed zero at $q=1$ to
$q_{L3z,+}=q_{L3z,-}=2$. 

As $w$ increases above 1 in the FCP interval, $q_{L3z,+}$ decreases
monotonically from 1, so that $q_{rmz}(L_1)$ remains the fixed zero at
$q=1$. This example shows that although individual zeros of $Ph(G,q,w)$ in the
$q$ plane are continuous functions of $w$, the maximal real zero $q_{rmz}(G)$
of $Ph(G,q,w)$ is a discontinuous function of $w$.  The reason for the
discontinuity in $q_{rmz}(G)$ is the confluence of two complex-conjugate roots
that come together and pinch the real axis (at $q=2$) as $w$ decreases through
$w=0$, abruptly producing a new maximal real root (of multiplicity 2).  Thus,
\beq
\lim_{w \to 0^+} q_{rmz}(L_3)=1 \ , 
\label{lim1}
\eeq
but $q_{rmz}(L_3)=2$ for $w=0$.  

In the FCP interval $w > 1$, the fixed zero at $q=1$ remains as $q_{mrz}(L_3)$,
since $q_{L3z,+}$ decreases below 1, while $q_{L3z,-}$ decreases below 0.
Indeed, the zero $q_{L3z,+}$ has a local maximum at $w=1$ and decreases
monotonically as $w$ increases above 1; $q_{L3z,+}$ passes through 0 as $w$
increases through the value $w=4$ and behaves asymptotically like $q_{L3z,+}
\sim -(1/2)(3-\sqrt{5} \ )w$ as $w \to \infty$.  The zero $q_{L3z,-}$ also
decreases monotonically from its value of 0 at $w=1$ through negative values as
$w$ increases above 1, and has the asymptotic behavior $q_{L3z,-} \sim -(1/2)(3
+ \sqrt{5} \ )w$ as $w \to \infty$.

As $w$ decreases through negative values, the double zero $q_{L3z,\pm}$ at
$q=2$ splits apart again. The zero $q_{L3z,+}$ increases monotonically as $w$
decreases through negative values, and grows asymptotically as $q_{L3z,+} \sim
(1/2)(3-\sqrt{5} \ )|w|$ as $w \to -\infty$.  The other zero, $q_{L3z,-}$ is a
non-monotonic function of $w$; it first decreases below 2, reaching a minimum
of $q_{Lz3,-}=9/5$ for $w=-1/5$ and then increases, passing through the value
$q_{Lz3,-}=2$ again as $w$ decreases through the value $w=-1$, and increasing
asymptotically as $q_{L3z,-} \sim (1/2)(3+\sqrt{5} \ )|w|$ as $w \to -\infty$.
As these examples show, there is somewhat complicated behavior of the
individual zeros as a function of $w$ for even a very simple graph such as
$L_3$, and this behavior is, understandably, more complicated for larger
graphs.

In this case and others one can avoid the discontinuous behavior of
$q_{rmz}(G)$ by restricting $w$ to the range $w > 0$. Doing this, we have
examples from these simple graphs that exhibit a continuous increase of
$q_{rmz}(G)$ as $w$ decreases from 1 to 0 and an example in which $q_{rmz}$ is
fixed, independent of $w$ in this DFCP interval $0<w<1$.  For $w$ in the FCP 
interval $w > 1$ we find cases where $q_{rmz}(G)$ decreases monotonically as
$w$ increases and also a case where $q_{rmz}(G)$ is fixed at 1.

Among the $Ph(C_n,q,w)$ polynomials, the case $n=2$ is the same as
$Ph(L_2,q,w)$ and for $n=3$, there are two $w$-independent zeros, at $q=1$ and
$q=2$, while the third occurs at $q=3(1-w)$.  As $w$ decreases from 1 to 0,
this third zero increases from 0 to 3. In the FCP range $w > 1$, this zero
decreases from 0 at $w=1$ to $-\infty$ as $w \to \infty$.  

A particularly simple case to discuss is that of complete graphs $K_n$.  For
these, as is evident from Eq. (\ref{phkn}), $Ph(K_n,q,w)$ has zeros in $q$ at
$q=1, \ 2,.., n-1$, and $q=n(1-w)$.

The zeros of $Ph(G,q,w)$ in $q$ as a function of $w$ for fixed $w$ satisfy
certain boundedness properties \cite{sokalbound}.  For $w=1$ and $w=0$, these
specialize to the bound for a (usual, unweighted) chromatic polynomial, namely
that if $q$ is a zero of $P(G,q)$, then $|q| \le a \Delta_{max}(G)$, where
$\Delta_{max}(G)$ denotes the maximal degree of the vertices in $G$ and $a
\simeq 7.964$ from Ref. \cite{sokalbound} (improved in
Ref. \cite{fp}). However, these zeros of $Ph(G,q,w)$ in $q$ are unbounded as
$|w| \to \infty$.  This is already evident in the simplest case of a single
vertex, for which $Ph(L_1,q,w)=q-1+w$, with a zero at $q=1-w$ with a magnitude
that goes to infinity as $|w| \to \infty$.  Some insight into the lack of
boundedness of the zeros of $Ph(G,q,w)$ for arbitrary $w$ can be gained by
examining the behavior of the factor $\prod_{i=1}^{k(G_i')}(q-1+w^{n(G_i')})$
in $Ph(G,q,w)$.  As $w \to \infty$, each of these factors goes to infinity also
unless $q$ behaves like $1-w^{n(G_i')}$, going to $-\infty$.  In principle, one
might imagine a cancellation occuring between different
$\prod_{i=1}^{k(G_i'}(q-1+w^{n(G_i')})$factors for different spanning subgraphs
$G' \subseteq G$, via different signs of the $(-1)^{e(G')}$ factor in
$Ph(G,q,w)$, nevertheless, this makes is understandable why, in the absence of
such cancellation, some zero(s) of $Ph(G,q,w)$ have magnitudes $|q| \to \infty$
as $|w| \to \infty$.  Note that this behavior cannot simply be attributed to
the frustration that occurs when $w$ gets large and positive, because it is
also true in the unphysical region for $w$ negative.

Although the positions of the zeros of $Ph(G,q,w)$ in $q$ are continuous
functions of $w$ and vice versa, this is not true of the asymptotic locus
${\cal B}$.  Indeed, we shall show below that for the $n \to \infty$ limit of
the circuit graph $C_n$, as $w$ decreases below one, regardless of how small
the magnitude of $1-w$ is, the part of ${\cal B}_q$ that crosses the real $q$
axis on the left jumps discontinuously to the right by one unit, so that this
crossing occurs at $q=1$ instead of at $q=0$.  (In contrast, the right-hand
part of ${\cal B}_q$ increases above 2 continuously as $w$ decreases below 1.)

\subsection{Zeros of $Ph(G,q,w)$ in $w$ as Functions of $q$} 

One may also study the zeros of $Ph(G,q,w)$ in $w$ as a function of $q$.  For
graphs $G$ containing at least one edge, $Ph(G,1,w)$ vanishes identically.  We
therefore take $q \ne 1$, although we shall consider the limit $q \to 1$ below.
We again consider some simple examples.  $Ph(L_1,q,w)=0$ for $w = 1-q$.  From
Eq. (\ref{phline2}), we find that $Ph(L_2,q,w)=0$ for $w=1-(q/2)$.  From
Eq. (\ref{phline3}), it follows that $Ph(L_3,q,w)=0$ at $w=w_{L3z,\pm}(q)$,
where
\beq
w_{L3z,\pm} = \frac{1}{2}\Big [ 5-3q \pm \sqrt{(q-1)(5q-9)} \ \Big ] \ . 
\label{wl3zpm}
\eeq
These roots are real for $q \le 1$ and $q \ge 9/5$, and form a
complex-conjugate pair for $1 < q < 9/5$.  For $q=0$, these roots are 1 and 4.
As $q$ increases from 0 to $1$, $w_{L3z,+}$ decreases monotonically from 4 to
1, while $w_{L3z,-}$ first decreases, reaching a minimum of 4/5 at $q=4/5$, and
then increases to 1.  Thus, at $q=1$, the roots coalesce to form a double root.
As $q$ increases above 1, they split apart to form a complex-conjugate pair,
with the magnitude of the imaginary part reaching a maximum at $q=7/5$, for
which $w_{L3z,\pm}=(1/5)(2 \pm \sqrt{5} \, i)$.  As $q$ increases further,
these roots move back to the real axis, coming together again at $w=-2/5$ as
$q$ increases through the value 9/5.  As $q$ increases further, $w_{L3z,-}$
decreases monotonically, while $w_{L3z,+}$ first increases to the value 0 at
$q=2$ and then decreases.  For the complete graph $K_n$, $Ph(K_n,q,w)$ has a
single zero in $w$ at
\beq
w_{Knz}=1 - \frac{q}{n} \ . 
\label{wkn}
\eeq

The zeros of $Ph(G,q,w)$ in $w$ as a function of $q$ are not, in general,
bounded, even for finite values of $q$.  This is a consequence of the fact that
the coefficient of the term in $Ph(G,q,w)$ of highest power of $w$ may vanish
as a function of $q$, in contrast to the fact that the coefficient of the term
in $Ph(G,q,w)$ of highest power in $q$ is a $w$-independent constant (namely,
1) and hence never vanishes. Generically, in the absence of cancellations, a
root of an algebraic equation diverges when the coefficient of the term of
highest degree vanishes.  This is evident in the quadratic equation $a w^2 + b
w + c = 0$, where $a$, $b$, and $c$ are functions of $q$ with no common
factors. Let us denote the set of values of $q$ where $a(q)=0$ as $\{q_0\}$.
One of the roots of this equation, $w = (2a)^{-1}(-b \pm \sqrt{b^2-4ac} \, )$,
diverges when $q$ approaches one of the values in the set $\{q_0\}$, since $a
\to 0$ in this limit.  A similar comment applies for algebraic equations of
higher degree.  Given that we have restricted ourselves, with no loss of
generality, to connected graphs $G$, these all contain at least one edge, 
except for the case of a single vertex.  Hence, for these graphs with at least
one edge, $Ph(G,q,w)$ contains the factor $(q-1)$.  It is thus convenient to
discuss the reduced coefficients $\bar \beta_{G,j}$ defined in
Eq. (\ref{phsumwbar}).  A zero of $Ph(G,q,w)$ in $w$ has a magnitude that
generically diverges when the coefficient of the term of highest degree in $w$,
namely the coefficient $\bar \beta_{G,d_w(G)}$, vanishes. 
This type of divergence can be absent if coefficient(s) $\bar \beta_{G,j}$ with
$j < d_w(G)$ also vanish sufficiently rapidly as $q$ approaches the value 
where $\bar\beta_{G,d_w(G)}$ vanishes. 

We give some simple examples of the divergences in zeros of $Ph(G,q,w)$ in $w$
as a function of $q$. For the line graph $L_4$, using our result in
Eq. (\ref{phline4}) above, we find that $Ph(L_4,q,w)$ has zeros in $w$ at
\beq
w_{L4z,1} = 2-q
\label{wl4z1}
\eeq
and
\beq
w_{L4z,2} = -\frac{(q-2)^2}{(3q-4)} \ . 
\label{wl4z2}
\eeq
As $q - (4/3) \to 0^\pm$, $w_{L4z,2} \to \mp \infty$.  This divergence is a
consequence of the fact that (i) the term in $Ph(L_4,q,w)$ of highest power in
$w$, namely $(q-1)(3q-4)w^2$, has a reduced coefficient
$\bar\beta_{L_4,2}=3q-4$ that vanishes at $q=4/3$ and (ii) the terms of lower
degree in $w$ do not vanish at $q=4/3$, as is clear from their
reduced coefficients $\bar\beta_{L_4,1}=2(q-2)(2q-3)$ and
$\bar\beta_{L_4,0}=(q-2)^3$.

Another  example is provided by the line graph $L_6$. From Eq. (\ref{phline6})
it follows that $Ph(L_6,q,w)$ has zeros in $w$ at
\beq
w_{L6z,1} = -\frac{(q-2)^2}{(2q-3)}
\label{wl6z1}
\eeq
and
\beq
w_{L6z,j} = \frac{(q-2)\Big [ 5-4q \pm \sqrt{8q^2-16q+9} \ \Big ]}{4(q-1)} \ ,
\quad j=2,3 \ , 
\label{wl6z23}
\eeq
where $j=2$ and $j=3$ correspond to the $+$ and $-$ signs, respectively.  As
$q-(3/2) \to 0^\pm$, $w_{L6z,1}(q) \to \mp \infty$, and as $q-1 \to 0^\pm$,
$w_{L6z,2} \to \mp \infty$ while $w_{L6z,3} \to 1$. The divergences in
$w_{L6z,1}$ and $w_{L6z,2}$ are consequences of the fact that (i) the term of
highest power in $w$ in $Ph(L_6,q,w)$, namely $2(q-1)^2(2q-3)w^3$, has a
reduced coefficient $\bar \beta_{L_6,3}=2(q-1)(2q-3)$ that vanishes at $q=1$
and $q=3/2$ and (ii) the terms of subleading degree in $w$ do not vanish at
either of these values of $q$, as is clear from their reduced coefficients
$\bar\beta_{L_6,2}= (q-2)(10q^2-28q+19)$, $\bar\beta_{L_6,1}= 2(3q-4)(q-2)^3$,
and $\bar\beta_{L_6,0}=(q-2)^5$.

For the $Y_5$ graph, using our result in Eq. (\ref{phy5}) for $Ph(Y_5,q,w)$, we
find that this polynomial has zeros in $w$ at
\beq
w_{Y5z,1}=2-q
\label{wy5z1}
\eeq
and
\beq
w_{Y5z,j} = \frac{\Big [ -4q^2 +13q-11 \pm (q-1)\sqrt{8q^2-28q+25} \ \Big ]}
{2(2q-3)} \ , \quad j=2,3 \ , 
\label{wy5z23}
\eeq
where $j=2$ and $j=3$ correspond to the $+$ and $-$ signs, respectively.  As
$q-(3/2) \to 0^\pm$, $w_{Y5z,3} \to \mp \infty$, while $w_{Y5z,2} \to 1/4$.
These divergences can easily be understood from the structure of the reduced
coefficients $\bar\beta_{Y_5,j}$ for the various powers of $w$ in
$Ph(Y_5,q,w)$, as discussed in general above.

As a last example, for the $IsoY_6$ graph, using our calculation in
Eq. (\ref{phisotree6}), we find that $Ph(IsoY_6,q,w)$ has a double zero at 
\beq
w_{IsoY6z,1}=2-q
\label{wisoy6z1}
\eeq
and two other zeros at 
\beq
w_{IsoY6z,j}=\frac{\Big [ -2q^2+6q-5 \pm (q-1)\sqrt{3q^2-10q+9} \ \Big ]}
{q-2} \ , \quad j=2,3 \ , 
\label{wisoy6zj}
\eeq
where $j=2,3$ for the $\pm$ signs, respectively, as before.  As $q-2 \to
0^\pm$, $w_{IsoY6z,3} \to \mp \infty$, while $w_{IsoY6,2} \to 0$.  

As these examples show, the zeros of $Ph(G,q,w)$ in $w$ can be unbounded as
functions of $q$.  The divergences in these zeros that we have found occur at
values of $q \ge 1$ (including the integers $q=1$ and $q=2$).  It is of
interest to determine a region in the $q$ plane for which the zeros of
$Ph(G,q,w)$ in $w$ are bounded, and we are studying this problem. Our results
show that such a region would have to exclude the real interval $1 \le q \le
2$.

\section{Quantities Defined in the Limit $n(G) \to \infty$}

\subsection{$\Phi$ Function} 

From the chromatic polynomial $P(G,q) \equiv Ph(G,q,1)$, one defines a
configurational degeneracy, which is the ground-state degeneracy, when 
viewing $P(G,q)$ as the partition function of the zero-temperature Potts 
antiferromagnet, 
\beq
W(\{G\},q) = \lim_{n \to \infty} P(G,q)^{1/n} \ , 
\label{w}
\eeq
where $n = n(G)$ and we use the symbol $\{G\}$ to denote the limit $n \to
\infty$ for a given family of graphs (and the symbol $W$ should not be confused
with the variable $w$).  In the present context, this $n \to \infty$ limit
corresponds to the limit of infinite length for a strip graph of fixed width
and some prescribed boundary conditions. The associated configurational entropy
per vertex (ground-state entropy per site of the Potts antiferromagnet) for 
$\{G\}$ is
\beq
S = k_B \ln W \ . 
\label{s}
\eeq
(There should not be any confusion between $W$ and $w$, which refer to
different quantities.)  The third law of thermodynamics states that the entropy
per site $S$ goes to zero as the temperature goes to zero.  However, there are
a number of exceptions to this law.  Elementary lower bounds on $W$ are $W \ge
(q-1)^{1/2}$ on a bipartite graph and $W \ge (q-2)^{1/3}$ on a tripartite
graph.  Hence $W > 1$, and $S > 0$ for (i) $q > 1$ and (ii) $q > 2$ on the $n
\to \infty$ limit of a (i) bipartite and (ii) tripartite graph,
respectively. In each of these cases, the third law is violated.  A well-known
violation in nature is water ice, which exhibits ground-state entropy
\cite{pauling}.  

In the present case with a nonzero external magnetic field $H
\ne 0$, we define an analogous quantity
\beq
\Phi(\{G\},q,w) = \lim_{n(G) \to \infty} Ph(G,q,w)^{1/n} \ . 
\label{phi}
\eeq
As before (cf. Eq. (1.9) of \cite{w} and Eq. (2.8) of \cite{a}), one must
take account of a noncommutativity of limits, namely the fact that for
certain special values of $q$, denoted $\{ q_s \}$, the limits $n \to \infty$
and $q \to q_s$ do not commute:
\beq
\lim_{n \to \infty} \lim_{q \to q_s} Ph(G,q,w)^{1/n} \ne
\lim_{q \to q_s} \lim_{n \to \infty} Ph(G,q,w)^{1/n} \ . 
\label{wnoncom}
\eeq
Because of these noncommutativities, the formal definition (\ref{phi}) is, in
general, insufficient to define $\Phi$ at these special points;
it is necessary to specify the order of the limits that one
uses in eq. (\ref{wnoncom}). This noncommutativity
also affects the resultant accumulation sets ${\cal B}$.  We have discussed
this in detail before in the case of the chromatic polynomial \cite{w} and
zero-field Potts model partition function \cite{a}. 
Modulo this subtlety, it follows from Eqs. (\ref{phw1}) and (\ref{phw0}) that 
\beq
\Phi(\{G\},q,1)=W(\{G\},q)
\label{phiw1}
\eeq
and 
\beq
\Phi(\{G\},q,0)=W(\{G\},q-1) \ . 
\label{phiw0}
\eeq

\subsection{Accumulation Locus ${\cal B}$ for Strip Graphs and $\Phi$ Function}

For a $n$-vertex graph $G$ in a recursive family of graphs such as strip
graphs, as $n \to \infty$, a subset of the zeros of $Ph(G,q,w)$ merge to form a
locus ${\cal B}$.  For fixed $w$, this is a locus ${\cal B}_q$ in the $q$
plane, while for fixed $q$, it is a locus ${\cal B}_w$ in the $w$ plane.  We
define a dominant (maximal) eigenvalue $\lambda_{max}$ as an eigenvalue whose
magnitude $|\lambda_{max}|$ is larger than or equal to the magnitudes of all
other eigenvalues.  From Eq. (\ref{phsum}), it follows that the zeros of
$Ph(G,q,w)$ can occur either as an isolated zero of a single dominant
eigenvalue or where two dominant eigenvalues are equal in magnitude.  The
continuous accumulation set of the zeros of $Ph(G,q,w)$ in a given variable,
denoted ${\cal B}$, is given generically by the solution set of the condition
of equality of dominant eigenvalues.  For real $w$, the coefficients of the
terms in $Ph(G,q,w)$ are also real (actually integers, although this is not
used here), and consequently, the set of zeros of $P(G,q,w)$ in the complex $q$
plane is invariant under complex conjugation $q \to q^*$.  {\it A fortiori},
the locus ${\cal B}_q$ is also invariant under this complex conjugation.  By
the same logic, for real $q$, the set of zeros of $Ph(G,q,w)$ in the complex
$w$ plane is invariant under the complex conjugation $w \to w^*$, and so is
their accumulation set ${\cal B}_w$. Because the $\lambda$'s are the same for
lattice strip graphs with cyclic and M\"obius boundary conditions, it follows
that in the $L_x \to \infty$ limit, the loci ${\cal B}$ are also the same.
With regard to the zeros of $Ph(G,q,w)$ in $w$ for fixed $q$, we note that
these are to be contrasted with the zeros of the Potts model partition function
$Z(G,q,v,w)$ in $w$ for fixed $q$ and $v \ne -1$, which have studied previously
in many works.

For the strip graphs of width $L_y$ considered here, 
\beq
\Phi(\{G\},q,w) = (\lambda_{max})^{1/L_y} \ . 
\label{wlam}
\eeq
As one moves across a locus ${\cal B}$, there is thus, generically, a switching
of dominant eigenvalues and an associated non-analyticity in $\Phi$.  
From Eq. (\ref{phw0}) we have 
\beq
\Phi(\{G\},q,0) = \Phi(\{G\},q-1,1) \ . 
\label{wwrel}
\eeq

\section{$\Phi$ Function and Accumulation Locus ${\cal B}$ for Line Graphs} 

\subsection{${\cal B}_q$}

Only $\lambda_{1,0,j}$, $j=1,2$ contribute to $Ph(L_n,q,w)$.  For $w=1$ we
encounter the noncommutativity of Eq. (\ref{wnoncom}).  If we first set $w=1$
and then vary $n$, we can use the fact that
$Ph(L_n,q,1)=P(L_n,q)=q(q-1)^{n-1}$, so that aside from the single zero at
$q=0$, the zeros accumulate at $q=1$ and ${\cal B}_q$ degenerates to this
single point.  If we choose the other order of limits, first taking $n \to
\infty$ and then $w \to 1$, then the locus ${\cal B}_q$ is the solution to the
equation $|\lambda_{1,0,1}|=|\lambda_{1,0,2}|$.  This equation is always
satisfied if $q=2$, so this point is on ${\cal B}_q$, which forms a
complex-conjugate arc with endpoints where $A_1=0$, where $A_1$ was given in
Eq. (\ref{a1}).  As $w \to 1$, these endpoints come together at $q=0$.  

For $w=0$, $Ph(L_n,q,0)=P(L_n,q-1)=(q-1)(q-2)^{n-1}$, so the locus ${\cal B}_q$
degenerates to the single point at $q=2$.  We proceed to consider $w \ne 0, \
1$.  Here, the equation $|\lambda_{1,0,1}|=|\lambda_{1,0,2}|$ determines the
locus ${\cal B}_q$. The DFCP interval $0 \le w \le 1$ is of particular
interest, since for this interval the weighted chromatic polynomial $Ph(G,q,w)$
interpolates between two chromatic polynomials; $Ph(G,q,1) = P(G,q)$ and
$Ph(G,q,0)=P(G,q-1)$.  We thus study ${\cal B}_q$ for this interval first.  In
this DFCP interval, ${\cal B}$ forms a (self-conjugate) arc passing through
$q=2$ and ending at the points where $A_1=0$, namely $q_{e,j}$, $j=1,2$,
\beq
q_{e,j} = 2 \Big [ 1-w \pm \sqrt{w(w-1)} \ \Big ] \ , \quad j=1,2  \ , 
\label{qej}
\eeq
where $j=1,2$ correspond to the $\pm$ signs, respectively.  Here the square
root in Eq. (\ref{qej}) is pure imaginary, so $q_{e,1} = q_{e,2}^*$ form a
complex-conjugate pair.  As $w$ decreases below 1, these arc endpoints move
away from the real axis near $q=0$.  They reach their maximal distance from the
real axis at $w=1/2$, where $q_{e,j}=1 \pm i$, and as $w$ decreases from
1/2 to 0, these arc endpoints come back toward this axis, finally reaching it
at $q=2$.

In the FCP interval $w > 1$, the locus ${\cal B}_q$ is a line segment whose
right and left ends occur at $q_{e,1}$ and $q_{e,2}$, respectively. As $w$
increases above $w=1$, the line segment extends outward from the point $q=0$. 
Some illustrative sets of values for these endpoints are 
(i) for $w=1.2$, $q_{e,1} \simeq 0.580$ and $q_{e,2} \simeq -1.380$, 
(ii) for $w=2$, $q_{e,1} \simeq 0.828$ and $q_{e,2} \simeq -4.828$, and
(iii) for $w=10$, $q_{e,1}=0.974$ and $q_{e,2}=-36.974$.   
As $w \to \infty$, the the right end of this line segment 
occurs asymptotically at 
\beq
q_{e,1} = 1-\frac{1}{4w}-\frac{1}{8w^2}-O\Big (\frac{1}{w^3}\Big ) \ , 
\label{qe1line}
\eeq
while the left end occurs at approximately 
\beq
q_{e,2} = -4w + 3 + \frac{1}{4w} + O \Big ( \frac{1}{w^2} \Big ) \ . 
\label{qe2line}
\eeq

As discussed above, the ranges of $w$ for these weighted coloring problems are
$0 \le w < 1$ and $w > 1$, respectively.  We may also consider an extension of
this range of values of $w$ in which $w$ becomes negative, although negative
values of $w$ do not correspond to a weighted graph coloring problem.  As $w$
decreases from 0 through negative values, ${\cal B}_q$ forms a line segment
that extends outward from the point $q=2$, with left and right ends at
$q_{e,1}$ and $q_{e,2}$, respectively.  As $w \to -\infty$, the left end
approaches $q=1$ from above, as $q_{e,1} \simeq 1+(4|w|)^{-1} + O(w^{-2})$,
while the right end goes to infinity, as $q_{e,2} \simeq 4|w| + 3 -
(4|w|)^{-1}$.  This locus ${\cal B}_q$ does not separate the complex $q$ into
any separate regions.  The dominant $\lambda$ is $\lambda_{1,0,1}$ and, 
denoting the formal limit of this family of line
graphs as $\lim_{n \to \infty} L_n = \{L\}$, the resultant $\Phi$ function is
\beq
\Phi(\{L\},q,w) = \lambda_{1,0,1} = 
\frac{1}{2}\Big [ q-2 + \sqrt{(q-2)^2+4(q-1)w} \ \Big ] \ . 
\label{philine}
\eeq
Here it is understood that one takes account of the branch cut associated
with the branch point singularity in the square root, so that at large negative
$q$, $|\lambda_{1,0,1}| \sim -q$.  For $ w \simeq 1$, this has the Taylor
series expansion
\beqs
\Phi(\{L\},q,w) & = & q-1 + \frac{[(q-1)(w-1)]}{q}-\frac{[(q-1)(w-1)]^2}{q^3} 
\cr\cr 
& + & \frac{2[(q-1)(w-1)]^3}{q^5} - O\Big ( (w-1)^4 \Big ) \quad {\rm as} 
\quad w \to 1 \ . 
\label{philine_taylor_w1}
\eeqs
For $w \simeq 0$, the Taylor series expansion for $\Phi(\{L\},q,w)$ is 
\beq
\Phi(\{L\},q,w) = (q-2) \Big [ 1 + z - z^2 + 2z^3 - O(z^4) \Big ] \quad 
{\rm as} \quad w \to 0  \ , 
\label{philine_taylor_w0}
\eeq
where here we use the compact notation 
\beq
z = \frac{(q-1)w}{(q-2)^2} \ . 
\label{zdef}
\eeq
As $|w| \to \infty$, $\Phi(\{L\},q,w)$ behaves asymptotically as 
\beq
\Phi(\{L\},q,w) \sim  \sqrt{(q-1)w} \ \Big [ 1 + \frac{q-2}{2\sqrt{(q-1)w}} 
+ O \Big ( \frac{1}{w} \Big ) \Big ] \quad {\rm as} \quad |w| \to \infty \ . 
\label{philine_large_w} 
\eeq
As $q \to \infty$, $\Phi(\{L\},q,w)$ behaves asymptotically as 
\beq
\Phi(\{L\},q,w) \sim  q-2+w - \frac{w(w-1)}{q} + \frac{2w(w-1)^2}{q^2} + 
O \Big ( \frac{1}{q^3} \Big ) \quad {\rm as} \quad q \to \infty  \ . 
\label{philine_large_q}
\eeq

\begin{figure}
\epsfxsize=3.5in
\epsffile{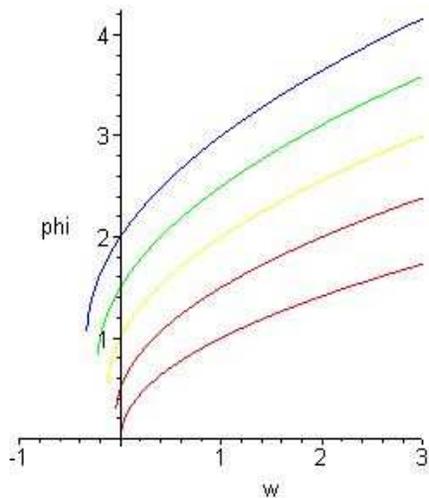}
\caption{Plot of $\Phi(\{L\},q,w)$ as a function of $w$ for the following
values of $q$, from bottom to top: $q=$ (a) 2 (b) 2.5 (c) 3 (d) 3.5 (e) 4.}
\label{philinewfig}
\end{figure}

In Fig. \ref{philinewfig} we show plots of $\Phi(\{L\},q,w)$ as a function of
$w$ for some representative values of $q$.  As is evident from Fig.
\ref{philinewfig}, for $q > 1$, $\Phi(\{L\},q,w)$ is a monotonically increasing
function of $w$ in the DFCP range $0 < w < 1$. This reflects the fact that it
is easier to carry out a proper $q$-coloring of the line graph as the weighting
factor $w$ increases from 0 to 1.  It is also easy to understand why
$\Phi(\{L\},q,w)$ is, for $q > 1$, a monotonically increasing function of $w$
in the FCP interval of real $w > 1$, with the leading asymptotic form given in
Eq. (\ref{philine_large_w}), $\Phi(\{L\},q,w) \sim \sqrt{(q-1)w}$. To see this,
let us start with finite $n$ and maximize the number of vertices assigned the
color 1, in order to maximize the power of $w$ in $\Phi(L_n,q,w)$. Since the
graph $L_n$ is bipartite, we can start at, say, the left end of the line graph
and assign the corresponding vertex, denoted vertex $i=1$ with this color 1.
Then the next vertex to the right, $i=2$, cannot be assigned this color, but
there are $q-1$ possibilities for its color.  The vertices $i=3, \ 5$, and so
forth for all odd-numbered vertices, are similarly assigned the color 1.  Each
of the even-numbered vertices can independently be assigned any of $q-1$
colors.  Taking into account all of these possible color assignments (or
equivalently, spin configurations, in the statistical mechanics context), it
follows that, if $n$ is even, then the dominant term in $\Phi(L_n,q,w)$ as $w
\to \infty$ for $q > 1$ is $Ph(L_n,q,w) = [(q-1)w]^{n/2}$ and if $n$ is odd,
then this dominant term is $Ph(L_n,q,w) = (q-1)^{(n-1)/2} \, w^{(n+1)/2}$.  In
either case, for $q > 1$, as $w \to \infty$, if one takes $n \to \infty$ and
calculates $\Phi(\{L\},q,w)$, one obtains, as the leading asymptotic
expression, $\Phi(\{L\},q,w) \sim \sqrt{(q-1)w}$.

Although negative $w$ is not associated with any coloring problem, we also show
in Fig. \ref{philinewfig} the extensions of the curves into the negative-$w$
region.  The function $A_1$
in the square root of $\Phi(\{L\},q,w)$ becomes negative, and hence
$\Phi(\{L\},q,w)$ becomes complex, for $w < w_z(q)$, where
\beq
w_z(q) = -\frac{(q-2)^2}{4(q-1)} \ . 
\label{wz}
\eeq
For example, $w_z(2)=0$, $w_z(3)=-1/8$, and $w_z(4)=-1/3$. We only plot each
curve for $w > w_z(q)$.  The values at $w=0$ and $w=1$ are
$\Phi(\{L\},0,w)=q-2$ and $\Phi(\{L\},1,w)=q-1$. 

\begin{figure}
\epsfxsize=3.5in
\epsffile{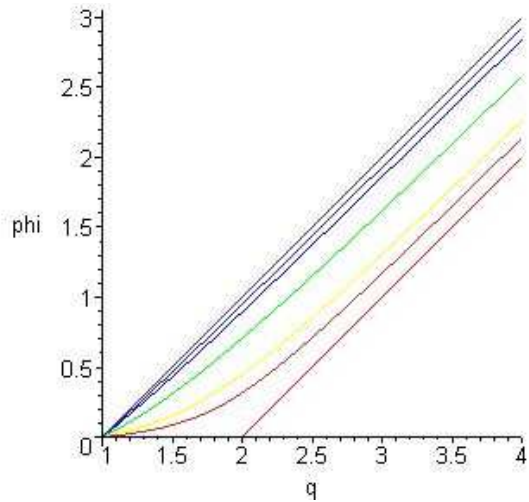}
\caption{Plot of $\Phi(\{L\},q,w)$ as a function of $q \ge 1$ for the following
values of $w$: \ $w=$ (a) 0.1 (b) 0.2 (c) 0.5 (d) 0.8 (e) 0.9. The plot
also shows $Ph(\{L\},q,0)=q-2$ and $Ph(\{L\},q,1)=q-1$. The curves for the $w$
values (a)-(e) are arranged from bottom to top, to the right of $q=1$, between 
these lines.}
\label{philineqfig}
\end{figure}
\begin{figure}
\epsfxsize=3.5in
\epsffile{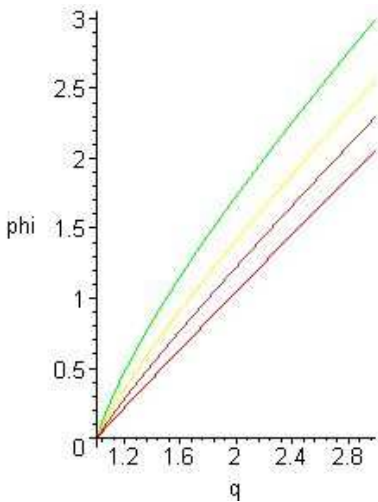}
\caption{Plot of $\Phi(\{L\},q,w)$ as a function of $q \ge 1$ for the following
values of $w$ in the FCP interval: $w=$ (a) 1.1 (b) 1.5 (c) 2 (d) 3. 
The curves for the $w$ values (a)-(d) are arranged from bottom to top.}
\label{philineq2fig}
\end{figure}

In Figs. \ref{philineqfig} and \ref{philineq2fig} we show $\Phi(\{ L \},q,w)$
as a function of $q$ for the range relevant for coloring, namely $q \ge 1$, and
a set of $w$ values in the DFCP interval $0 < w < 1$ and the FCP interval $w >
1$, respectively.  One sees that for fixed $q > 1$, $\Phi(\{L\},q,w)$ is a
monotonically increasing function of $w$ for $w > 0$. Given its definition as
the $1/n$'thd power of the weighted chromatic polynomial $Ph(G,q,w)$ as $n \to
\infty$ and the fact that $Ph(G,q,w)$ provides a measure
of the ease of carrying out a weighted proper $q$-coloring of the
graph $G$, it follows that where $\Phi(\{G\},q,w)$ is meaningful for such
colorings, it is non-negative.  Hence the contination of the line that passes
through $q=2$ to negative values of $\Phi$ is not relevant for graph coloring.

These calculations with line graphs show a number of general features of the
weighted chromatic polynomial $Ph(G,q,w)$ and the associated limiting function
$\Phi(G,q,w)$.  They suggest the following conjectured generalization for
weighted colorings of families of strip graphs $G$ of regular lattices
$\Lambda$: in the $n \to \infty$ limit of $G$, denoted $\{G\}$.  Let the
chromatic number of the lattice $\Lambda$ be indicated as $\chi(\Lambda)$.  For
a bipartite lattice such as a line graph $L_n$, a circuit graph $C_n$ with even
$n$, or a square, cubic, or body-centered cubic lattice,
$\chi(\Lambda_{bip.})=2$.  For a triangular lattice, $\chi(tri)=3$, etc.
Assume $q \ge \chi(\Lambda)$.  Then (i) for fixed $w > 0$, $\Phi(\{G\},q,w)$ is
a monotonically increasing function of $q$ and (ii) for fixed $q$,
$\Phi(\{G\},q,w)$ is a monotonically increasing function of $w$ for $ w > 0$.
The generalization (i) is understandable since $\Phi(G,q,w)$ is a measure of
the number of weighted proper $q$-coloring of $G$, and this should increase if
there are more colors, i.e., if $q$ increases.  For the DFCP interval $0 < w <
1$, the generalization (ii) follows because increasing $w$ in this interval
removes the penalty factor for coloring a vertex with one of the colors and
hence clearly makes it easier to perform a proper $q$-coloring of the vertices
of $G$.  In the FCP interval $w > 1$, the generalization (ii) is rendered
plausible because for sufficiently large $q$, one can analyze the dominant
terms contributing to $\Phi(G,q,w)$, and these arise from maximizing the number
of vertices that can be assigned the color 1, subject to the proper
$q$-coloring condition, and then enumerating the possible color assignments for
the other vertices.  We remark that for these monotonicity properties, it is
important that $q \ge \chi(G)$.  As an example of how the behavior differs when
$q < \chi(G)$, consider the weighted chromatic polynomial of the complete
graph, $K_n$, given in Eq. (\ref{phkn}).  Recall that $\chi(K_n)=n$.  Let us
take $n=4$ for definiteness, so that $Ph(K_4,q,w) = (q-1)(q-2)(q-3)[q+4(w-1)]$.
This is a monotonically increasing function of $q$ if $q > 3$, but not for
smaller values of $q$.  Moreover, say we keep $w$ arbitrary but choose the
illustrative value $q=5/2$, whence $Ph(K_4,5/2,w) = (3/16)(3-8w)$. This is not
an increasing function of $w$ in the DFCP or FCP intervals.

\subsection{${\cal B}_w$}

One can also study ${\cal B}_w$ as a function of $q$.  We find that ${\cal
B}_w$ is the semi-infinite real line segment
\beq
{\cal B}_w: \quad w < w_z(q) \quad {\rm for} \quad \{L\} \ , 
\label{bwline}
\eeq
where $w_z(q)$ was given in Eq. (\ref{wz}). For the range of $q$ relevant for
weighted graph coloring, namely $q > 1$, $w_z(q) \le 0$.

\section{$\Phi$ Function and Accumulation Locus ${\cal B}_q$ for Circuit 
Graphs} 

Here we discuss the accumulation set ${\cal B}$ of the zeros of $Ph(G,q,w)$ 
as $n \to \infty$ for the family of circuit graphs, $C_n$ (this limit is
denoted $\{C\}$).  The results depend on $w$, so we discuss various intervals
of $w$ in turn.

\subsection{ $w=1$} 

We begin by briefly reviewing the results for the case unweighted case $w=1$,
i.e., for the chromatic polynomial $P(C_n,q)$, given in Eq. (\ref{pcn}). The
resultant locus ${\cal B}_q$ in the limit $n \to \infty$ is the unit circle
\beq
{\cal B}_q: \quad |q-1|=1 \quad {\rm for} \quad w=1 \ . 
\label{bqw1}
\eeq
This crosses the real axis at the points
\beq
q_{cr,1}=0  \ , \quad q_{cr,2} = 2 \quad {\rm for} \quad w=1 \ . 
\label{chrom_qc_w1}
\eeq
Following our earlier notation \cite{w,wcyl,s4}, we denote the maximal point at
which ${\cal B}_q$ intersects the real $q$ axis for the $n \to \infty$ limit of
a given family of graphs, $\{G\}$, as $q_c(\{G\})$.  Thus, here, $q_c=2$. 

Indeed, all of the complex zeros lie exactly on this unit circle \cite{wc}.
The boundary ${\cal B}_q$ separates the $q$ plane into two regions, in which
$W(q)$ has different analytic forms.  Outside of the circle $|q-1|=1$, the
dominant $\lambda$ is $\lambda_{1,0,1}=q-1$, while inside of this circle, the
dominant $\lambda$ is $\lambda_{1,1}=-1$, so
\beqs
W(q) & = & q-1 \quad {\rm for} \quad |q-1| > 1 \cr\cr
|W(q)| & =& 1 \quad {\rm for} \quad |q-1| < 1 \ . 
\label{wqq1}
\eeqs

\subsection{ $0 \le w < 1$}

As $w$ decreases below 1 in the interval $0 \le w < 1$, the boundary ${\cal
B}_q$ continues to be a simple closed curve separating the $q$ plane into two
regions.  However, there is a discontinuous change in the form of this
boundary.  On the left, the point at which the locus ${\cal B}_q$ crosses the
real $q$ axis jumps from $q=0$ for $w=1$ to
\beq
q_{cr,1} = 1  \quad {\rm for} \ \ 0 \le w < 1 \ . 
\label{chrom_qc_win01}
\eeq
Associated with this, the left part of the boundary ${\cal B}_q$ changes
discontinuously from a section of a circle to an involuted cusp with its tip at
$q=1$. (There may also be discrete zero(s) to the left of $q=1$.) 
Thus, as one moves along the curve forming ${\cal B}_q$ upward from the
point $q=1$ where it intersects the real axis in the cusp, this curve moves to
the upper left, finally curving around to go upward, and then over to the
right.  The behavior of the boundary ${\cal B}_q$ on the right
side is continuous as $w$ deviates from 1; this part of ${\cal B}_q$ crosses
the real axis at
\beq
q_{cr,2} = q_c = \frac{w+3}{w+1}  \quad {\rm for} \quad \{G\}=\{C\} \ . 
\label{qcr2}
\eeq
This point $q_c$ increases monotonically from $q_c=2$ for $w=1$ to $q_c=3$ as
$w$ decreases from 1 to 0. We denote as $R_1$ the region that includes the real
interval $q > q_c$ and the part of the complex $q$ plane analytically connected
to it, which is the region outside of the closed curve formed by ${\cal B}_q$,
For $w$ only slightly less than unity, ${\cal B}_q$ has the form of a lima
bean, with its concave part facing left and its convex part facing right.  As
$w$ decreases through this interval $0 \le w < 1$, the bulbous parts of ${\cal
B}_q$ on the upper left and lower left disappear, and eventually, as $w$
decreases toward 0, the locus ${\cal B}_q$ becomes the circular locus
(\ref{bqw1}) with $q$ replaced by $q-1$ (in accordance with Eq. (\ref{phw0})),
i.e., unit circle whose center is shifted horizontally by one unit to the right
in the $q$ plane:
\beq
{\cal B}_q: \quad |q-2|=1 \quad {\rm for} \ \ w=0 \ . 
\label{bqw0}
\eeq

In region $R_1$, 
\beq
\Phi(\{C\},q,w) =\Phi(\{L\},q,w) \quad {\rm for} \quad q \in R_1 \ , 
\label{wr1cn}
\eeq
where $\Phi(\{L\},q,w)$ was given above in Eq. (\ref{philine}). 
In region $R_2$ forming the interior of the closed curve ${\cal B}_q$,
\beq
|\Phi(\{C\},q,w)| = 1 \quad {\rm for} \quad q \in R_2 \ . 
\label{wr2cn}
\eeq
Thus, a plot of $\Phi(\{C\},q,w)$ as a function of $q$ for $w \in (0,1)$ is
similar to the plot of $\Phi(\{L\},q,w)$ given in Fig. \ref{philineqfig} but
with the difference that the curve extends only down to the value $q=q_c$ given
in Eq. (\ref{qcr2}), where $\Phi(\{L\},q_c,w)=\Phi(\{C\},q_c,w)=1$, and for $1
\le q \le q_c$, $\Phi(\{C\},q,w)$ has unit magnitude (in region $R_2$, while
$\Phi(\{L\},q,w)$ continues downward, reaching zero as $q \to 1$. 

\subsubsection{ $w > 1$}

For the range $w > 1$, the locus ${\cal B}_q$ contains a line segment 
on the real axis, whose left end occurs at $q_{e,2}$ (cf. Eq. (\ref{qej})). 
This is a solution of the condition that $A_1=0$, where $A_1$ is the function
in the square root in Eq. (\ref{a1}).  Along this line segment, $A_1 < 0$, so
that the square root in Eq. (\ref{lam1d0j}) is pure imaginary; hence, the
$\lambda_{1,0,j}$ with $j=1,2$ are complex conjugates of each other. They are
also larger than $|\lambda_{1,1}|=1$ and hence this line segment is on ${\cal
  B}_q$. As one moves to the right along the real axis on this line segment,
when one comes to the intermediate point $q_{int}$, given by 
\beq
q_{int} = 1 - \frac{1}{w} \ , 
\label{qint}
\eeq
the equal magnitudes of $\lambda_{1,0,j}$, $j=1,2$ decrease through the value
1, so this is the point at which this line segment on ${\cal B}_1$
terminates.  For the present range $w > 1$, the point $q=0$ is always on this
line segment, since 
\beq
\lambda_{1,0,j}(q=0) = -1 \pm i\sqrt{w-1} \ , \quad j=1,2 \ , 
\label{lam10jq0}
\eeq
and these are dominant over $\lambda_{1,1}=-1$, since $|\lambda_{1,0,j}| =
1+|w-1|$ for $j=1,2$.  In the real interval
\beq
q_{int} \le q \le 1 \ , 
\label{qint1}
\eeq
and in the region of the complex $q$ plane analytically connected with it,
$\lambda_{1,1}=-1$ is dominant.  We denote this region as $R_3$, and 
\beq
|\Phi(\{C\},q,w)| = 1 \quad {\rm for} \quad q \in R_3 \ . 
\label{phicnr3}
\eeq
Thus, although the square root in the $\lambda_{1,0,j}$ is imaginary for the
full interval $q_{e,2} \le q \le q_{e,1}$, this does not affect ${\cal B}_q$
for $q > q_{int}$.

At $q=1$, $\lambda_{1,0,2}$ becomes degenerate with $\lambda_{1,1}$,
so ${\cal B}_q$ crosses the real $q$ axis at this point.  The eigenvalue
$\lambda_{1,0,2}$ is dominant in the interval
\beq
1 \le q \le 2 
\label{q12}
\eeq
and the region of the $q$ plane analytically connected with it.  We denote this
region as $R_2$, and obtain 
\beq
\Phi(\{C\},q,w) = \frac{1}{2}\Big [ q-2 - \sqrt{(q-2)^2 + 4(q-1)w} \ \Big ]
\quad {\rm for} \quad q \in R_2 \ . 
\label{phicnr2}
\eeq
At the point 
\beq
q = q_c =2
\label{qc}
\eeq
$\lambda_{1,0,1}=-\lambda_{1,0,2}=\sqrt{w}$, and both are dominant over 
$\lambda_{1,1}=-1$, since $w > 1$. Hence, ${\cal B}_q$ crosses the
real $q$ axis again at this point, and this is the maximal real value of $q$
where ${\cal B}_q$ crosses the real axis.  We denote the real interval $q > 2$
and the region of the complex $q$ plane analytically connected with it as
region $R_1$.  In this region, since $\lambda_{1,0,1}$ is dominant,
\beq
\Phi(\{C\},q,w) = \frac{1}{2}\Big [ q-2 + \sqrt{(q-2)^2 + 4(q-1)w} \ \Big ]
\quad {\rm for} \quad q \in R_1 \ . 
\label{phicnr1}
\eeq
Thus, for $w > 1$, the locus ${\cal B}_q$ for the $n \to \infty$ limit of 
$Ph(C_n,q,w)$ separates the $q$ plane into three regions.

\begin{figure}
\epsfxsize=3.5in
\epsffile{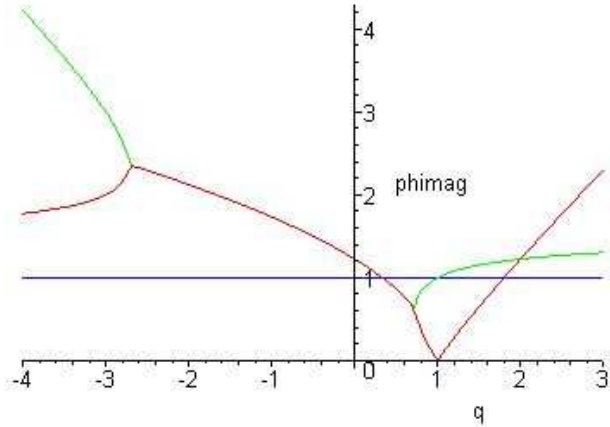}
\caption{Plot of (a) $|\lambda_{1,0,1}|$, (b) $|\lambda_{1,0,2}|$, and (c)
$|\lambda_{1,1}|=1$ for $Ph(C_n,q,w)$ with the illustrative value $w = 3/2$.
The values of these magnitudes and the corresponding order of the curves from
top to bottom are (i) $a > b > c$ for $q > 2$; (ii) $b > a > c$ for $q \in
(1,2)$, (iii) $c > b > a$ for $q_{int} < q < 1$, (iv) $a=b>c$ for $q_{e,2} < q
< q_{int}$, (v) $a > b > c$ for $q < q_{e,2}$.  For $w=3/2$, $q_{int}=1/3$ and
$q_{e,2} \simeq -2.732$. See text for further discussion.}
\label{lambdamagfig}
\end{figure}
In Fig. \ref{lambdamagfig} we plot $|\lambda_{1,0,1}|$, $|\lambda_{1,0,2}|$,
and $|\lambda_{1,1}|=1$ as functions of $q$ for the illustrative value
$w=3/2$. For this value, $q_{int}=1/3$ and $q_{e,2}=-(1+\sqrt{3} \ ) \simeq
-2.732$.  As before, the $\lambda$ of dominant magnitude determines
$\Phi(\{C\},q,w)$. For $q > 2$, $\lambda_{1,0,1}$ is dominant; for $1 \le q \le
2$, $\lambda_{1,0,2}$ is dominant; for $1/3 \le q \le 1$, $\lambda_{1,1}$ is
dominant, and for $q < q_{e,2}$, $\lambda_{1,0,1}$ is again dominant (taking
into account the branch cut in the definition of the sign of the square root).
On the interval $q_{e,2} \le q \le q_{int}$, $|\lambda_{1,0,1}|$ and
$|\lambda_{1,0,2}|$ are equal in magnitude and dominant.  This plot shows the
intersections of the curves (or line) where there is a degeneracy of dominant
$\lambda$'s at $q=2$, $q=1$, and along the interval $q_{e,2} \le q_{int}$.

\subsection{ $w < 0$}

As $w$ decreases below 0 through the range $-1/3 < w < 0$, ${\cal B}_q$ forms a
closed curve that crosses the real axis at the fixed point $q=q_{cr,1}=1$ and
at the $w$-dependent point $q=q_{cr,2}$, which increases as $w$ becomes more
negative.  The shape of ${\cal B}_q$ changes as $w$ becomes more negative in
this interval, becoming a teardrop with its broadly rounded end on the left and
its sharper end on the right.  As $w$ decreases through the value $w=-1/3$, a
line segment appears, and for $w < -1/3$, the rightmost part of the locus
${\cal B}$ is comprised by this line segment, which extends from $q=q_{int}$ in
Eq. (\ref{qint}) to $q_{e,1}$ in Eq. (\ref{qej}).  Thus, for $w < -1/3$,
the locus ${\cal B}$ consists of a teardrop-shaped curve crossing the real axis
on the left at $q=1$ and on the right at $q=q_{int}$ (with the bulbous part of
the teardrop on the left and the sharp part on the right), together with a line
segment extending over the interval $q_{int} \le q \le q_{e,1}$.  At
$w=-1/3$,
\beq
(\lambda_{1,0.j})_{w=-1/3} = \frac{1}{2}\Big [ (q-2) \pm 
\sqrt{(q-4)(q-(4/3))} \ \Big ] \ . 
\label{lam10j_weqminus1over3}
\eeq
For this value $w=-1/3$, and $q=4$, the pair $\lambda_{1,0,j}$, $j=1,2$, are
equal to each other and are also equal to the magnitude $|\lambda_{1,1}|=1$. As
$w \to -\infty$, $q_{int} \to 1^+$, so the teardrop curve contracts to a point
at $q=1$ while the right-hand endpoint of the line segment at $q_{e,1}$
approaches infinity like $q_{e,1} \sim 4|w|$.

One can also calculate the loci ${\cal B}_q$ for the $n \to \infty$
limits of other families of graphs.  However, our discussion for the $n \to
\infty$ limits of line graphs $L_n$ and circuit graphs $C_n$ already exhibit 
a number of salient features of these loci.

\section{ Locus ${\cal B}_w$ for Circuit Graphs}

Here we discuss the locus ${\cal B}_w$ as a function of $q$ for the $n \to
\infty$ limit of the circuit graph $C_n$.  First, we note that if $q=1$ or
$q=2$, then we encounter the noncommutativity (\ref{wnoncom}).  For $q=1$,
this is evident from Eq. (\ref{qminus1factor}), according to which if we set
$q=1$ first and then take $n \to \infty$, the problem is trivial, since
$Ph(C_n,1,w)$ vanishes identically. For $q=2$, the noncommutativity is evident
from our general result in Eq.  (\ref{phcnq2}), because the coefficient of the
$(\lambda_{1,1})^n$ term in $Ph(C_n,q,w)$ vanishes if $q=2$. If we first take
$n \to \infty$ and then set $q=2$, we have $\lambda_{1,0,j}=\pm \sqrt{w}$, so
that the locus ${\cal B}_w$ is the union of the semi-infinite real line
segments $w > 1$ and $w < -1$.  If, on the other hand, we first set $q=2$, and
then vary $n$, our result (\ref{phcnq2}) shows that the limit of $Ph(C_n,2,w)$
as $n \to \infty$ does not exist, since $Ph(C_n,2,w)$ is alternatively zero for
odd $n$ and $2w^{n/2}$ for even $n$.  If we restrict to odd $n$, then the
problem of the zeros of $Ph(C_n,2,w)$ is trivial since the function itself
vanishes identically, while if we restrict to even $n$, then the locus ${\cal
B}_w$ degenerates to a point at $w=0$, correswponding to the zero at this point
with multiplicity $n/2$.

From Eq. (\ref{phq0}) we know that for $q=0$, $Ph(G,0,w)$ contains a factor of
$(w-1)$ for an arbitrary graph $G$, and this is true, in particular, for
$G=C_n$.  The other zeros occur at real values $w > 1$.  For this value $q=0$,
it follows that $\lambda_{1,0,j}=-1 \pm \sqrt{1-w}$, and these are dominant
$\lambda$'s if $w \ne 1$, so ${\cal B}_w$ is the semi-infinite real line
segment $w \ge 1$.  For $q \ne 0, \ 1, \ 2$ we typically find that the locus
${\cal B}_w$ may consist of the union of a (self-conjugate) loop and a line
segment.  Details depend on the specific value of $q$.

\section{ ${\cal B}$ for Wheel Graphs}

We have also calculated the locus ${\cal B}$ for wheel graphs.  We denote the
$n \to \infty$ limit of the graph $Wh_n$ as $\{Wh\}$. For $w=1$, 
${\cal B}_q$ is the unit circle $|q-2|=1$, which crosses the real axis at 
$q=1$ and $q_c=3$ and separates the $q$ plane into two regions.  In the region
with $|q-2|>1$, i.e., the region exterior to this circle, $\Phi(\{Wh\},q,1) =
W(\{Wh\},q) = q-2$. In the for which $|q-2| < 1$, i.e., the region interior 
to the circle, $|\Phi(\{Wh\},q,1)| = |W(\{Wh\},q)|=1$.  

As $w$ decreases from 1 in the DFCP interval, the boundary ${\cal B}_q$
continues to form a closed curve separating the $q$ plane into two regions, as
it did for $w=1$, but there is a discontinuous jump in the crossing point on
the left, from $q=1$ to $q=2$.  This is similar to what we found for the $n \to
\infty$ of the circuit graph, where the jump in the crossing point on the left
was from $q=0$ to $q=1$.  The crossing point on the right is 
\beq
q_c = \frac{2(w+2)}{w+1} \quad {\rm for} \quad \{G\} = \{C\} 
\label{qcwheel}
\eeq
Region $R_1$ includes the real interval $q > q_c$ and the portion of the
complex $q$ plane analytically connected with this interval, and thus lying
outside of the boundary ${\cal B}_q$.  Region $R_2$ occupies the portion of the
$q$ plane inside of the boundary ${\cal B}_q$.  The dominant $\lambda$ in
region $R_1$ is $\lambda_{Wh,+}$, while the dominant $\lambda$ in $R_2$ is
equal to $-1$.  The point $q_c$ occurs where these are degenerate in
magnitude. Given this and the relation (\ref{lamrel_wheelcircuit}), it follows
that $q_c$ for $\{G\} = \{Wh\}$ is related to $q_c$ for $\{G\}=\{C\}$, given in
Eq. (\ref{qcr2}), by replacing $q$ by $q-1$.  That is, if one replaces $q_c$ on
the left-hand side of Eq. (\ref{qcr2}) by $q_c-1$ and solves for the new $q_c$,
one obtains Eq. (\ref{qcwheel}).  This is in accord with the fact that the
proper $q$-coloring of the wheel graph with $q$ colors is closely related to
the proper coloring of the circuit graph with $q-1$ colors.  As $w$ decreases
from 1 to 0 in the DFCP interval the $q_c$ in Eq. (\ref{qcwheel}) increases
continuously from 3 to 4.  One can also analyze other ranges of $w$ and the
locus ${\cal B}_w$ in a similar manner.

\section{Some Observations and Conjectures}

\subsection{Sign Alternation of Successive Terms in $Ph(G,q,w)$}

One can write the chromatic polynomial $P(G,q)$ of a graph as 
\beq
P(G,q) = \sum_{j=0}^{n-1} \alpha_{G,n-j}\, q^{n-j} \ , 
\label{pform}
\eeq
where, without loss of generality, we take $G$ to be connected. 
The signs of the coefficients $\alpha_{G,n-j}$ alternate: 
\beq
{\rm sgn}(\alpha_{G,n-j}) = (-1)^j \ , \quad 0 \le j \le n-1 \ . 
\label{aj}
\eeq
This is proved by iterated application of the deletion-contraction theorem.
Since the weighted chromatic polynomial $Ph(G,q,w)$ does not, in general, obey
a deletion-contraction theorem, except for the values $w=1$ and $w=0$ for which
it reduces to a chromatic polynomial (see Eqs. (\ref{phw1}) and (\ref{phw0})),
one does not expect the coefficients $\alpha_{G,n-j}(w)$ in $Ph(G,q,w)$ to have
this sign-alternation property, and they do not.  However, from our analysis of
weighted coloring polynomials for several families of graphs, we have noticed
that for a restricted range of $w$, namely $0 \le w \le 1$, this sign
alternation again holds, namely ${\rm sgn}(\alpha_{G,n-j}(w)) = (-1)^j$ for $0
\le j \le n-1$.  For $j=n$, namely for the $q^0$ term in $Ph(G,q,w)$, the sign
alternation also holds for $0 \le w < 1$; here the coefficient $\alpha_{G,0}$
contains the factor $(w-1)$ and hence vanishes at $w=1$.  We also find that
this sign alternation property holds, as far as we have checked it, for real
negative $w$.  It is of interest to investigate whether this sign alternation
property for $w < 1$ holds on other families of graphs.  We are currently
continuing with this investigation.

Related to this, it is of interest to study where the coefficients
$\alpha_{G,n-j}(w)$ vanish in the complex $w$ plane.  From our calculation of
weighted chromatic polynomials for line and circuit graphs $L_n$ and $C_n$, we
have observed that for the graphs we have considered, 
the coefficients $\alpha_{L_n,n-j}$ and $\alpha_{C_n,n-j}$
for $1 \le j \le n-1$ have zeros in the real interval $w > 1$, while for the 
coefficients $\alpha_{L_n,0}$ and $\alpha_{C_n,0}$ have, in addition to the
always-present zero at $w=1$ (recall Eqs. (\ref{phq0}) and (\ref{alphaq0})),
the other zeros, if any, again occur in the real interval $w > 1$.  In
contrast, we find that for other graphs, the coefficients $\alpha_{G,n-j}$ may
have complex-conjugate pairs of zeros.  For example, in $Ph(S_4,q,w)$, the
coefficient of the $q$ term, $\alpha_{S_4,1}=w^3-9w^2+27w-20$, has zeros
at $w \simeq 1.087$ and $w \simeq 3.9565 \pm 1.6566i$, and the coefficient of
the $q^0$ term, $\alpha_{S_4,0}=-(w-1)(w^2-5w+8)$, has zeros at $w=1$ and $w = 
(1/2)(5 \pm \sqrt{7} \, i)$.  Similarly, coefficients of star graphs $S_n$ for
larger $n$ include cases having complex-conjugate pairs of zeros in the $w$
plane.  

\subsection{Generalized Unimodal Conjecture}

From his study of chromatic polynomials, R. Read observed that the magnitudes
of the coefficients of successive powers of $q^{n-j}$, $0 \le j \le n-k(G)$ in
a chromatic polynomial satisfy a unimodal property \cite{rtrev}.  That is, the
magnitudes of these coefficients get successively larger and larger, and then
smaller and smaller, as $j$ increases from 0 to $n-k(G)$.  There is thus a
unique maximal-magnitude coefficient, or two successive coefficients whose
magnitudes are equal. From our calculations of weighted chromatic polynomials
for a number of families of graphs, we have observed that in the interval $0
\le w \le 1$ this property continues to hold.  We therefore state the following
conjecture: {\it Conject.} Let $Ph(G,q,w)$ be written as in
Eq. (\ref{phsum}). Then for real $w$ in the interval $0 \le w \le 1$, the
quantities $(-1)^j\alpha_{G,n-j}(w)$, $0 \le j \le n$, are positive and satisfy
the unimodal property, i.e., $(-1)^j\alpha_{G,n-j}(w)$ get progressively larger
and larger, and a maximal value is reached for a given $j$, or for two 
successive $j$ values, and then the quantities  $(-1)^j\alpha_{G,n-j}(w)$ get
pogressively smaller, as $j$ increases from 0 to $n$.

\section{Some Generalizations}

Although we have focused in this paper on the proper $q$-coloring of vertices
with one color given a disfavored or favored weighting, we discuss some
generalized weighted coloring problems in this section. We first present an
extension of Eq. (\ref{clusterw}) to the most general case of different fields
$H_p$, $p=1,...,q$, and hence different weighting factors, for each of the $q$
different colors.  The generalization to multiple fields corresponding to
different spin values in the Potts model was noted, e.g., in \cite{sf} and
\cite{kj} and more recently in \cite{sokalbound}. The Hamiltonian for this case
is
\beq
{\cal H} = -J \sum_{\langle i j \rangle} \delta_{\sigma_i, \sigma_j}
- \sum_{p=1}^q \Big [ H_p \sum_\ell \delta_{\sigma_\ell,p} \Big ] \ .  
\label{genham}
\eeq
Let us define
\beq
h_p = \beta H_p \ , \quad w_p = e^{h_p} \quad {\rm for} \quad 1 \le p \le q 
\label{ws}
\eeq
and denote the set of $w_p$, $p=1,...,q$ as $\{w\}$.  The partition function is
a function of $q$, $v$, and $\{w\}$, and hence we write it as $Z(G,q,v,\{w\})$.
For $v=-1$, the resultant generalized weighted chromatic polynomial is
$Ph(G,q,\{w\})=Z(G,q,-1,\{w\})$.

We have derived the following generalization of the Wu formula for this case 
(where, as before, $G'$ is a spanning subgraph of $G$):
\beq
Z(G,q,v,\{w\}) = \sum_{G' \subseteq G} v^{e(G')} \
\prod_{i=1}^{k(G')} \Big (\sum_{p=1}^q w_p^{n(G'_i)} \Big ) \ .
\label{clusterwp}
\eeq
This is proved as follows. The spins in each component $G_i'$ of
$G'$ are connected by edges, so they all have the same value, and there are $q$
possibilities for this value.  For a given spanning subgraph $G'$, the
weighting factor is the product $\prod_{i=1}^{k(G')} \Big ( \sum_{p=1}^q
w_p^{n(G'_i)} \Big )$.  This subgraph thus contributes a term $v^{e(G')}
\prod_{i=1}^{k(G')} \Big ( \sum_{p=1}^q w_p^{n(G'_i)} \Big )$ to $Z$.  Summing
over all spanning subgraphs $G'$ then yields the result (\ref{clusterwp}). 
$\Box$  

The resultant spanning graph formula for the generalized weighted chromatic
polynomial $Ph(G,q,\{w\})$ is obtained by evaluating Eq. (\ref{clusterwp}) at
$v=-1$:
\beq
Ph(G,q,\{w\}) = \sum_{G' \subseteq G} (-1)^{e(G')} \
\prod_{i=1}^{k(G')} \Big (\sum_{p=1}^q w_p^{n(G'_i)} \Big ) \ . 
\label{phclusterwp}
\eeq
Note that some $w_p$'s may disfavor certain color(s), i.e., $0 \le w_p < 1$,
while others may favor other color(s), $w_{p'} > 1$.  Note also that in the
general situation with different $H_p$, $p=1,...,q$, the dependence of
$Ph(G,q,\{w\})$ on $q$ appears via the $w_p$, $p=1,...,q$ rather than via a
polynomial dependence on the variable $q$.

Let us illustrate this generalization for the case where a set of $s$ colors
is subject to a given (disfavored or favored) weighting, i.e. (where without
loss of generality, we label these $s$ colors as $1,\ 2,..., s$)
\beq
H_p = \cases{ H \ne 0 & for $1 \le p \le s$ \cr
            0 & for $s+1 \le p \le q$ }
\label{hvalues}
\eeq
so that 
\beq
w_p = \cases{ w \ne 1 & for $1 \le p \le s$ \cr
              1       & for $s+1 \le p \le q$ } \ . 
\label{wvalues}
\eeq
Then, with $Z(G,q,v,\{w\})$ written compactly as $Z(G,q,s,v,w)$ in an
evident notation, Eq. (\ref{clusterwp}) takes the form
\beq
Z(G,q,s,v,w) = \sum_{G' \subseteq G} v^{e(G')} \
\prod_{i=1}^{k(G')} \Big (q-s + s w^{n(G'_i)} \Big ) 
\label{clusterwgen}
\eeq
(To avoid awkward notation, we use the same symbol $Z$ for the Potts
model partition function with the various sets of arguments, $Z(G,q,v)$,
$Z(G,q,v,w)$, $Z(G,q,v,\{w\})$, and $Z(G,q,s,v,w)$.)  Writing the weighted
chromatic polynomial $Ph(G,q,v,\{w\})$ in the same notation as $Ph(G,q,s,w)$,
we have
\beq
Ph(G,q,s,w) = Z(G,q,-1,s,w) \ . 
\label{phzgen}
\eeq
With Yan Xu at Stony Brook, a study of the properties of the generalized
weighted chromatic polynomial $Ph(G,q,s,w)$ has been carried out, and the
results will be reported elsewhere.  We note here that since $s$ only appears
in Eq. (\ref{clusterwgen}) in the combination
\beq
\prod_{i=1}^{k(G')}\Big (q-s+s w^{n(G'_i)} \Big ) = 
\prod_{i=1}^{k(G')}\Big (q+s(w-1)\sum_{r=0}^{n(G'_i)-1}w^r \Big ) \ , 
\label{auxeqs}
\eeq
it follows that $Z(G,q,s,w)$ and $Ph(G,q,s,w)$ can equivalently be written as
polynomials in the variables $q$, $v$, 
\beq
t=s(w-1) \ , 
\label{tvar}
\eeq
and $w$. We mention the following general relations involving 
$Z(G,q,s,v,w)$ and $Ph(G,q,s,w)$ that hold for $t=0$ :
\beq
Z(G,q,0,v,w)=Z(G,q,s,v,1)=Z(G,q,v) \ , 
\label{zs0_zsw1}
\eeq
\beq
Ph(G,q,0,w)=Ph(G,q,s,1)=P(G,q) \ , 
\label{phs0_phsw1}
\eeq
and the general relations that hold for $w=0$: 
\beq
Z(G,q,s,v,0)=Z(G,q-s,v) \ , 
\label{zsw0}
\eeq
and
\beq
Ph(G,q,s,0)=P(G,q-s) \ . 
\label{phsw0}
\eeq
From Eqs. (\ref{zs0_zsw1}) and (\ref{phs0_phsw1}), it is clear that for 
$s=0$, the $Z$ and $Ph$ polynomials reduce to their zero-field forms.  A
similar reduction to a factor times the zero-field forms occurs if $s=q$:
\beq
Z(G,q,q,v,w)=w^{n(G)}Z(G,q,v)
\label{zsq}
\eeq
and
\beq
Ph(G,q,q,w)=w^{n(G)}P(G,q) \ . 
\label{phsq}
\eeq
Hence, we are primarily interested in the values $s=1,..,q-1$. Assuming that
$G$ contains at least one edge, then, if $q=1$, it is impossible to satisfy the
proper $q$-coloring constraint, so $Ph(G,1,s,w)=0$.  This vanishing does not,
in general, result as a consequence of an explicit $(q-1)$ factor, unless $s=0$
or $s=1$.  Instead, when one sets $q=1$ in $Ph(G,q,s,w)$, one obtains a
polynomial with a factor of $s(s-1)(w-1)^2$.  Since $s$ is a non-negative
integer bounded above by $q$, the condition that $q=1$ implies that $s$ is
either 0 or 1, and hence this factor must vanish, yielding the necessary result
that $Ph(G,1,s,w)=0$.

A different type of generalization is to have the spin-spin couplings depend on
the edges of the graph $G$, so they would be of the form $J_{ij} \equiv J_e$,
where $i$ and $j$ denote adjacent vertices of $G$ connected by the edge
$e$. The study of spin models with spin-spin couplings that are different for
different lattice directions goes back to the early decades of the twentieth
century, reflecting the fact that there are often anisotropies in real magnetic
substances. In the 1960's and 1970's, anisotropic spin-spin couplings were
studied to investigate how, for various ferromagnetic and antiferromagnetic
combinations on different lattices, they could affect critical behavior
\cite{fisherrmp}. In the 1970's and later the further generalization to
spin-spin couplings that depend on each edge was studied in connection with
disordered materials and the question of how such disorder changed the critical
behavior \cite{sg}; and discussions of edge-dependent $J_{ij}$ continue
\cite{sokalbound}.  The Hamiltonian for the general Potts model for this case,
including the full set of $q$ different external fields, is
\beq
{\cal H} = - \sum_{\langle i j \rangle} J_{ij} \delta_{\sigma_i, \sigma_j}
- \sum_{p=1}^q \Big [ H_p \sum_\ell \delta_{\sigma_\ell,p} \Big ] \ . 
\label{genhamje}
\eeq
Let us define 
\beq
K_{ij} = \beta J_{ij}, \quad v_{ij} = e^{K_{ij}} - 1
\label{vij}
\eeq
and denote the set of all $v_{ij}$ as $\{v\}$.  The partition function is then
$Z(G,q,\{v\},\{w\})$.  We give the following general formula for this partition
function, where again $G'=(V,E')$ is a spanning subgraph of $G$:
\beq
Z(G,q,\{v\},\{w\}) = \sum_{G' \subseteq G} \ \Big [ \prod_{e \in E'} v_e 
\Big ] \
\Big [\prod_{i=1}^{k(G')} \Big ( \sum_{p=1}^q w_p^{n(G'_i)} \Big ) \Big ] \ . 
\label{clusterwpje}
\eeq
For the weighted proper $q$-coloring problem, $v_e = -1 \ \forall \ e \in E$,
so this generalization reduces to Eq. (\ref{phclusterwp}). Having derived and
presented spanning graph formulas for the Potts model partition function and
weighted chromatic polynomial for these generalized cases, we focus henceforth
on the simple case where only one color is subject to the (disfavored or
favored) weighting.  Note, as before, that in the general situation, the
dependence of $Z(G,q,\{v\},\{w\})$ on $q$ appears via the $w_p$, $p=1,...,q$
rather than via a polynomial dependence on the variable $q$.
We illustrate these generalizations with the circuit graph $G=C_3$.  Let us
define $\eta_r = \sum_{p=1}^q w_p^r$. Then from Eq. (\ref{clusterwpje}) we have
\beq
Z(C_3,q,\{v\},\{w\}) = \eta_1^3+(v_{12}+v_{23}+v_{31})\eta_2 \eta_1 
+ \Big [(v_{12}v_{23}+v_{23}v_{31}+v_{31}v_{12})+v_{12}v_{23}v_{31} \Big ] 
\eta_3
\label{zc3mostgen}
\eeq
and, setting $v_e=-1$ for all of the edges in $C_3$, we obtain 
$Ph(C_3,q,\{w\}) = \eta_1^3-3\eta_2\eta_1+2\eta_3$.

Yet another generalization is to make the sets of colors that one chooses from
to assign to each vertex depend on the vertex. With the weighting, this defines
a new weighted list-coloring problem. A practical realization of this problem
is the allocation of frequencies to radio broadcasting or wireless mobile
communication transmitters where each individual transmitter has its own set of
available frequencies, no adjacent transmitters should use the same frequency,
and there are various disfavored and/or favored frequencies.  For a graph
$G=(V,E)$, we denote the list of available colors for a given vertex as
$\{c_i\}$, where $i=1,...,n(G)$, and we denote the set of all color lists with
the symbol $\{ \{ c \} \} \equiv \{ \{c_1\},...,\{c_n\} \}$.  We define the
associated partition function as $Z(G,\{ \{c\} \}, \{v\}, \{w\})=
\sum_{\{\sigma_i\}}\exp(-\beta {\cal H})$, with
\beq
{\cal H} = - \sum_{\langle i j \rangle} J_{ij}\delta_{\sigma_i, \sigma_j}
- \sum_{p=1}^q \Big [ H_p \sum_\ell \delta_{\sigma_\ell,p} \Big ] \ , 
\label{hamlist}
\eeq
where $\sigma_i$ takes on values in the list $\{c_i\}$. As before, one may
consider special cases of this weighted list coloring problem in which, e.g.,
the $J_{ij}$ are constants, independent of the edge joining the vertices $i$
and $j$, and/or where the $H_p$ are of the simple form (\ref{hvalues}), etc.
As a simple example, we again take the circuit graph $G=C_3$ and choose the
available color lists for each vertex as $\{c_1\}=(1,2)$, $\{c_2\}=(2,3)$,
$\{c_3\}=(1,3)$.  The generalized weighted chromatic polynomial for this
weighted list coloring problem would then be $2w_1w_2w_3$ corresponding to the
color assignments $(1,2,3)$ and $(2,3,1)$ to vertices $i=1, \ 2, \ 3$.  In
passing, we recall the case where there are no external fields or corresponding
weightings, i.e., $H_p=0$, so $w_p=1$ for all $p=1,...,q$.  This is the usual
unweighted list coloring problem, as reviewed, e.g., in
Ref. \cite{woodall}. For example, for the case $G=C_3$ with the color lists
given above, the list chromatic polynomial is 2.  In contrast, for $G=C_3$ with
color lists $\{c_1\}=(1,2,3)$, $\{c_2\}=(1,2)$, and $\{c_3\}=(1)$ there is only
one proper coloring, namely the color assignment $(3,2,1)$ to vertices $1, \ 2,
\ 3$, so the list coloring polynomial is equal to 1.  

\section{Conclusions}

In this paper we have studied proper $q$-colorings of the vertices of a graph
with a weighting factor $w$ that either disfavors or favors a given color. In
particular, we have analyzed a weighted chromatic polynomial $Ph(G,q,w)$
associated with this problem, which generalizes the chromatic polynomial
$P(G,q)$.  Since $Ph(G,q,w)$ can be obtained as a special limit of the Potts
model partition function in an external magnetic field, its study represents a
fruitful confluence of statistical mechanics and mathematical graph theory.  We
have found a number of interesting properties of this weighted chromatic
polynomial. Among others, we have shown how it encodes more information about
the graph $G$, as shown by the fact that it is able to distinguish between
certain graphs that yield the same chromatic polynomial.  We have given
formulas for $Ph(G,q,w)$ for various families of graphs $G$, including line
graphs, star graphs, complete graphs, and cyclic lattice strip graphs.  For $w
\in (0,1)$, $Ph(G,q,w)$ effectively interpolates between $P(G,q)$ and
$P(G,q-1)$.  Using our results, we have discussed the zeros of $Ph(G,q,w)$ in
the $q$ and $w$ planes and their accumulation sets in the limit of infinitely
many vertices of $G$.  Finally, we have mentioned some observations,
conjectures, and related weighted graph-coloring problems.  There is ample
motivation for further research on this very interesting subject.

\begin{acknowledgments}

This research was partly supported by the grants Taiwan
NSC-97-2112-M-006-007-MY3 and NSC-98-2119-M-002-001 (S.-C.C.) and
U.S. NSF-PHY-06-53342 (R.S.).

\end{acknowledgments}

\newpage

\appendix 

\section{Tables on Structural Properties} 

For comparison with our new results for $n_{Ph}(L_y,d)$ and $N_{Ph,L_y}$, we
list here corresponding tables for the following numbers for cyclic strips of
the square (sq), triangular (tri), and honeycomb (hc) lattices $\Lambda$: 
(i)   $n_P(L_y,d)$ and their sums, $N_{P,L_y,\lambda}$ for the chromatic
polynomial with $h=0$,
(ii)  $n_Z(L_y,d)$ and their sums, $N_{Z,L_y,\lambda}$ for the Potts model
partition function with $h=0$, and
(iii) $n_{Zh}(L_y,d)$ and their sums, $N_{Zh,L_y,\lambda}$ for the Potts model
partition function with $h \ne 0$ \ \cite{hl,zth}. 

\begin{table}
\caption{\footnotesize{Table of numbers $n_P(L_y,d)$ and their sums, 
$N_{P,L_y,\lambda}$ for the chromatic polynomial of cyclic strips of the
lattice $\Lambda$ (sq, tri, hc) with $h=0$. Blank entries are zero. See text
for further discussion.}}
\begin{center}
\begin{tabular}{|c|c|c|c|c|c|c|c|c|c|c||}
\hline\hline
$L_y \ \downarrow$ \ \ $d \ \rightarrow$
   & 0 & 1   & 2   & 3   & 4   & 5  & 6  & 7 & 8 & $N_{P,L_y,\lambda}$
\\ \hline\hline
1  & 1   & 1   &     &     &     &    &    &       &       &    2   \\ \hline
2  & 1   & 2   & 1   &     &     &    &    &       &       &    4   \\ \hline
3  & 2   & 4   & 3   & 1   &     &    &    &       &       &   10   \\ \hline
4  & 4   & 9   & 8   & 4   & 1   &    &    &       &       &   26   \\ \hline
5  & 9   & 21  & 21  & 13  & 5   & 1  &    &       &       &   70   \\ \hline
6  & 21  & 51  & 55  & 39  & 19  & 6  & 1  &       &       &  192   \\ \hline
7  & 51  & 127 & 145 & 113 & 64  & 26 & 7  & \ 1 \ &       &  534   \\ \hline
8  & 127 & 323 & 385 & 322 & 203 & 97 & 34 & \ 8 \ & \ 1 \ & 1500   \\ 
\hline\hline
\end{tabular}
\end{center}
\label{nptable}
\end{table}

\begin{table}
\caption{\footnotesize{Table of numbers $n_Z(L_y,d)$ and their sums,
$N_{Z,G,\lambda}$, for the Potts model partition function on cyclic strips of
    the lattice $\Lambda$ (sq, tri, hc) with $h=0$. Blank entries are zero.  
See text for further discussion.}}
\begin{center}
\begin{tabular}{|c|c|c|c|c|c|c|c|c|c|c|}
\hline\hline
$L_y \ \downarrow$ \ \ $d \ \rightarrow$
   & 0 & 1   & 2   & 3   & 4   & 5  & 6  & 7 & 8 & $N_{Z,L_y,\lambda}$
\\ \hline\hline
1  & 1   & 1   &     &     &     &    &    &   &       &     2    \\ \hline
2  & 2   & 3   & 1   &     &     &    &    &   &       &     6    \\ \hline
3  & 5   & 9   & 5   & 1   &     &    &    &   &       &    20    \\ \hline
4  & 14  & 28  & 20  & 7   & 1   &    &    &   &       &    70    \\ \hline
5  & 42  & 90  & 75  & 35  & 9   & 1  &    &   &       &   252    \\ \hline
6  & 132 & 297 & 275 & 154 & 54  & 11 & 1  &   &       &   924    \\ \hline
7  & 429 & 1001& 1001& 637 & 273 & 77 & 13 & 1 &       &  3432    \\ \hline
8  & 1430& 3432& 3640& 2548& 1260& 440& 104& 15& \ 1 \ & 12870    \\ 
\hline\hline
\end{tabular}
\end{center}
\label{nztable}
\end{table}

\begin{table}
\caption{\footnotesize{Table of numbers $n_{Zh}(L_y,d)$ and their sums,
$N_{Zh,L_y}$ for the Potts model partition function on strips of the lattice
    $\Lambda$ (sq, tri, hc) with $h \ne 0$. Blank entries are zero. See text
    for further discussion.}}
\begin{center}
\begin{tabular}{|c|c|c|c|c|c|c|c|c|c|c|}
\hline\hline
$L_y  \ \backslash \ d$
   & 0    & 1    & 2    & 3    & 4    & 5   & 6   & 7  & 8 & $N_{Zh,L_y}$ \\
\hline\hline
1  & 2    & 1    &      &      &      &     &     &    &   &     3   \\ \hline
2  & 5    & 5    & 1    &      &      &     &     &    &   &    11   \\ \hline
3  & 15   & 21   & 8    & 1    &      &     &     &    &   &    45   \\ \hline
4  & 51   & 86   & 46   & 11   & 1    &     &     &    &   &   195   \\ \hline
5  & 188  & 355  & 235  & 80   & 14   & 1   &     &    &   &   873   \\ \hline
6  & 731  & 1488 & 1140 & 489  & 123  & 17  & 1   &    &   &  3989   \\ \hline
7  & 2950 & 6335 & 5397 & 2730 & 875  & 175 & 20  & 1  &   & 18483   \\ \hline
8  & 12235& 27352& 25256& 14462& 5530 & 1420& 236 & 23 & \ 1 \ & 86515   \\
\hline\hline
\end{tabular}
\end{center}
\label{nzhtable}
\end{table}

\newpage

\section{$Ph(G,q,w)$ for Tree Graphs $G$}

\subsection{$n=5$ Vertices}

There are three tree graphs with $n=5$ vertices, as shown in Fig.
Fig. \ref{tree_n5fig}: (i) the line graph $L_5$, (ii) the graph $Y_5$, and
(iii) the star graph $S_5$.  The weighted chromatic polynomials for these are
\beqs
Ph(L_5,q,w) & = & (q-1)\Big [ q^4+(5w-8)q^3+3(2w^2-9w+8)q^2 \cr\cr
            & +& (w-2)(w^2-16w+16)q-(w-1)(w^2-12w+16) \Big ] \cr\cr
            & = & q^5-(9-5w)q^4+2(w-4)(3w-4)q^3-(-w^3+24w^2-75w+56)q^2 \cr\cr
            & + & (-2w^3+31w^2-76w+48)q-(1-w)(w^2-12w+16) 
\label{phline5}
\eeqs
\beqs
Ph(Y_5,q,w) &=& (q-1)(q+w-2)\Big [q^3+2(2w-3)q^2+(2w^2-13w+12)q-(w-1)(3w-8) 
\Big ] \cr\cr
            &=& q^5-(9-5w)q^4+2(w-4)(3w-4)q^3-2(-w^3+13w^2-38w+28)q^2 \cr\cr
            &+& (-5w^3+37w^2-79w+48)q-(w-1)(w-2)(8-3w)
\label{phy5}
\eeqs
\beqs
Ph(S_5,q,w) & = & (q-1)\Big [ q^4+(5w-8)q^3+3(2w^2-9w+8)q^2 \cr\cr
            & + & (4w^3-24w^2+51w-32)q +(w-1)(w^3-7w^2+17w-16) \Big ] \cr\cr
            & = & q^5-(9-5w)q^4+2(w-4)(3w-4)q^3-2(-2w^3+15w^2-39w+28)q^2 \cr\cr
            & + & (w^4-12w^3+48w^2-84w+48)q-(w-1)(w^3-7w^2+17w-16) 
\label{phstar5}
\eeqs

\subsection{$Ph(G,q,w)$ for Tree Graphs with $n=6$ Vertices} 

\begin{figure}[htbp]
\unitlength 0.9mm
\begin{picture}(170,10)
\multiput(0,0)(10,0){6}{\circle*{1.6}}
\put(0,0){\line(1,0){50}}
\put(25,-5){\makebox(0,0){$L_6$}}
\multiput(70,0)(10,0){5}{\circle*{1.6}}
\put(80,10){\circle*{1.6}}
\put(70,0){\line(1,0){40}}
\put(80,0){\line(0,1){10}}
\put(90,-5){\makebox(0,0){$Y_6$}}
\multiput(130,0)(10,0){5}{\circle*{1.6}}
\put(150,10){\circle*{1.6}}
\put(130,0){\line(1,0){40}}
\put(150,0){\line(0,1){10}}
\put(150,-5){\makebox(0,0){$iso - Y_6$}}
\end{picture}

\vspace*{25mm}

\begin{picture}(170,20)
\multiput(0,0)(10,0){4}{\circle*{1.6}}
\multiput(10,10)(10,0){2}{\circle*{1.6}}
\put(0,0){\line(1,0){30}}
\multiput(10,0)(10,0){2}{\line(0,1){10}}
\put(15,-5){\makebox(0,0){$H_6$}}
\multiput(50,10)(10,0){4}{\circle*{1.6}}
\multiput(60,0)(0,20){2}{\circle*{1.6}}
\put(50,10){\line(1,0){30}}
\put(60,0){\line(0,1){20}}
\put(65,-5){\makebox(0,0){$Cr_6$}}
\multiput(100,10)(10,0){3}{\circle*{1.6}}
\multiput(105,0)(10,0){2}{\circle*{1.6}}
\put(110,20){\circle*{1.6}}
\put(100,10){\line(1,0){20}}
\put(110,10){\line(0,1){10}}
\put(110,10){\line(1,-2){5}}
\put(110,10){\line(-1,-2){5}}
\put(110,-5){\makebox(0,0){$S_6$}}
\end{picture}
\vspace*{5mm}
\caption{\footnotesize{Tree graphs with $n=6$.}}
\label{tree_n6fig}
\end{figure}
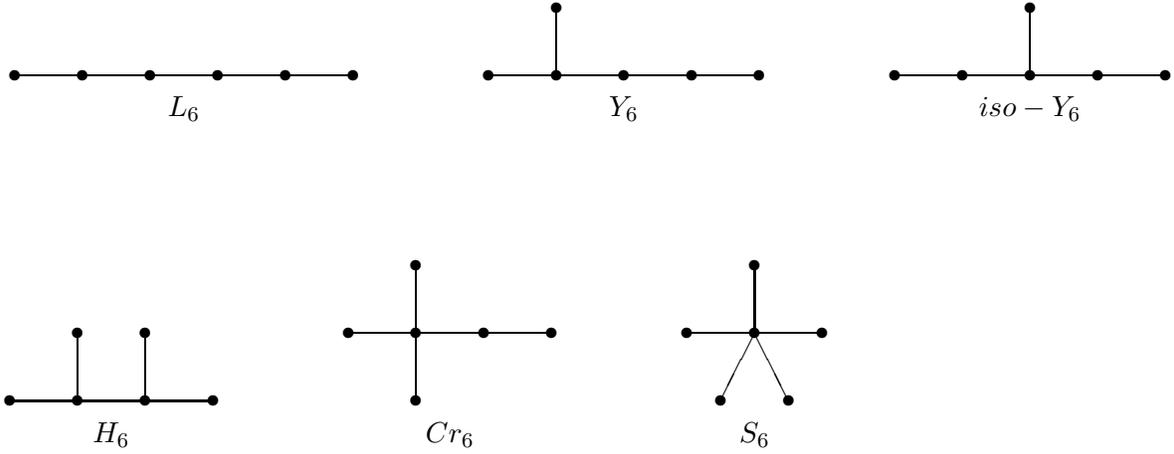

There are six tree graphs with $n=6$ vertices, as shown in
Fig. \ref{tree_n6fig}: (i) the line graph $L_6$, (ii) the graph $Y_6$, (iii)
the graph with a branch in the middle of the line, denoted $iso-Y_6$ (iv) a
graph with two branches, denoted $H_6$, (v) a graph forming a cross, denoted
$Cr_6$, and (vi) the star graph $S_6$.  (Again, we order these in terms of
graphs with increasing maximal vertex degree $\Delta$; one has $\Delta=2, \ 3,
\ 3, \ 3, \ 4, \ 5$ for graphs (i)-(vi), respectively.  In chemical
nomenclature, five of these graphs correspond to the carbon backbones of the
following alkanes: (i) n-hexane, (ii) 2-methylpentane, (iii) 3-methylpentane,
(iv) 2,3-dimethylbutane, and (v) 2,2-dimethylbutane.  The graph $S_6$ has no
carbon-atom correspondence, since the central vertex has degree 5. We find
\beqs
Ph(L_6,q,w) & = & (q-1)\Big [ q^2+2(w-2)q-3w+4 \Big ] \Big [ q^3 + 2(2w-3)q^2
  \cr\cr
& + & (2w^2-13w+12)q-2(w-1)(w-4) \Big ] \cr\cr
& = & q^6-(11-6w)q^5+10(w^2-5w+5)q^4-2(-2w^3+29w^2-82w+60)q^3 \cr\cr
& + & (-14w^3+123w^2-264w+160)q^2-(-16w^3+113w^2-208w+112)q \cr\cr
& + & 2(1-w)(4-3w)(4-w)
\label{phline6}
\eeqs
\beqs
Ph(Y_6,q,w) & = & (q-1)\Big [ q^5+2(3w-5)q^4+2(5w^2-22w+20)q^3 \cr\cr
            & + & (5w^3-50w^2+121w-80)q^2 +(w^4-16w^3+84w^2-148w+80)q \cr\cr
            & - & (w-1)(w-4)(w^2-7w+8) \Big ] \cr\cr
            & = & q^6 -(11-6w)q^5 + 10(w^2-5w+5)q^4 -5(-w^3+12w^2-33w+24)q^3
\cr\cr
            & + & (w^4-21w^3+134w^2-269w+160)q^2 \cr\cr
            & - & (2w^4-28w^3+131w^2-216w+112)q
\cr\cr      & + & (w-1)(w-4)(w^2-7w+8)
\label{phy6}
\eeqs
\beqs
Ph(IsoY_6,q,w) & = & (q-1)(q-2+w)\Big [ q^4+(5w-8)q^3+(w-4)(5w-6)q^2 \cr\cr
               & + & (-14w^2+45w-32)q+2(w-1)(5w-8) \Big ] \cr\cr
           & = & q^6 -(11-6w)q^5+10(w^2-5w+5)q^4-5(-w^3+12w^2-33w+24)q^3 \cr\cr
           & + & (-19w^3+133w^2-269w+160)q^2-(-24w^3+129w^2-216w+112)q \cr\cr
           & + & 2(w-1)(w-2)(8-5w)
\label{phisotree6}
\eeqs
\beqs
Ph(H_6,q,w) & = & (q-1)(q-2+w)^2 \Big [ q^3+2(2w-3)q^2+(w^2-12w+12)q
-2(w-1)(w-4) \Big ] \cr\cr
             & = & q^6-(11-6w)q^5+10(w^2-5w+5)q^4-2(-3w^3+31w^2-83w+60)q^3
\cr\cr       & + & (w^4-26w^3+144w^2-274w+160)q^2-(w-2)(3w^3-32w^2+84w-56)q
\cr\cr       & + & 2(w-1)(w-2)^2(w-4)
\label{phyy6}
\eeqs
\beqs
Ph(Cr_6,q,w) & = & (q-1)(q-2+w) \Big [ q^4+(5w-8)q^3+(w-4)(5w-6)q^2 \cr\cr
            & + & (2w^3-18w^2+47w-32)q-(w-1)(3w^2-13w+16) \Big ] \cr\cr
           & = & q^6-(11-6w)q^5+10(w^2-5w+5)q^4-(-7w^3+64w^2-167w+120)q^3\cr\cr
           & + & (2w^4-32w^3+153w^2-278w+160)q^2
      -(5w^4-47w^3+160w^2-229w+112)q \cr\cr
           & + & (w-1)(w-2)(3w^2-13w+16)
\label{phcross6}
\eeqs
\beqs
Ph(S_6,q,w) & = & (q-1)\Big [ q^5+2(3w-5)q^4+2(5w^2-22w+20)q^3 \cr\cr
            & + & 2(5w^3-30w^2+63w-40)q^2+(5w^4-40w^3+120w^2-164w+80)q \cr\cr
            & + & (w-1)(w^4-9w^3+31w^2-49w+32) \Big ] \cr\cr
            & = & q^6 -(11-6w)q^5+10(w^2-5w+5)q^4-10(-w^3+7w^2-17w+12)q^3\cr\cr
            & + & 5(w^4-10w^3+36w^2-58w+32)q^2 \cr\cr
            & - & (-w^5+15w^4-80w^3+200w^2-245w+112)q \cr\cr
            & + & (1-w)(w^4-9w^3+31w^2-49w+32)
\label{phstar6}
\eeqs

Among other things, these calculations can be used to characterize further the
way in which the weighted chromatic polynomial is able to distinguish between
graphs that yield the same chromatic polynomial.  As discussed in the text, all
tree graphs with a given number $n$ of vertices yield the same chromatic
polynomial, $P(G_{tree,n},q)=q(q-1)^{n-1}$ (and, indeed, also the same Tutte
polynomial $T(G_{tree_n},x,y) = x^{n-1}$).  Using our results above, we
calculate the following differences in weighted chromatic polynomials, relative
to $Ph(S_6,q,w)$, for definiteness, from which all other differences can be
obtained:
\beq
Ph(S_6,q,w)-Ph(L_6,q,w)=w(w-1)^2(q-1)\Big [ (3q+w)(2q+w)-20q-8w+17 \Big ] 
\label{phstar6_minus_phline6}
\eeq
\beq
Ph(S_6,q,w)-Ph(Y_6,q,w) = w(w-1)^2(q-1)(5q^2+4wq+w^2-16q-7w+13)
\label{phstar6_minus_phy6}
\eeq
\beq
Ph(S_6,q,w)-Ph(IsoY_6,q,w)=w(w-1)^2(q-1)\Big [5q^2+5wq+w^2-16q-8w+13\Big ]
\label{phstar6_minus_phisotree6}
\eeq
\beq
Ph(S_6,q,w)-Ph(H_6,q,w)=w(w-1)^2(q-1)(2q-3+w)^2
\label{phstar6_minus_phyy6}
\eeq
\beq
Ph(S_6,q,w)-Ph(Cr_6,q,w)=w(w-1)^2(q-1)\Big [3q^2+3wq+w^2-9q-5w+7 \Big ]
\label{phstar6_minus_cross6}
\eeq
We thus find that the weighted chromatic polynomials for all of the different
$n$-vertex tree graphs of a given $n$ are, in general, different from each
other, although they coincide for $w=1$ and $w=0$, where they reduce to
chromatic polynomials, and for $q=1$, where they all vanish.

\vfill
\eject
\end{document}